\documentclass[12pt]{elsart}
\usepackage{epsfig,graphics}

\newcommand{\tr}{\rm tr \,}

\newcommand{\one}{\bf 1}

\newcommand{\barqslash}{\FMslash {\bar q}}

\newcommand{\wslash}{\FMslash w}

\newcommand{\qslash}{\FMslash q}

\newcommand{\TableHeader}[0]{\begin{tabular}{|c|c|c|c|c|c|} \hline (I,S)  &  Kanal &  Lage &  |g|  & Lage  &|g| \\}
\def\rescale{\fontsize{6}{1}}

\begin{document}
\begin{frontmatter}
\title{D-wave baryon resonances with charm\\ from coupled-channel dynamics}
\author[GSI]{J. Hofmann}
\author[GSI]{and M.F.M. Lutz}
\address[GSI]{Gesellschaft f\"ur Schwerionenforschung (GSI)\\
Planck Str. 1, 64291 Darmstadt, Germany}
\begin{abstract}
Identifying the zero-range exchange of vector mesons as the
driving force for the s-wave scattering of pseudo-scalar mesons
off the baryon  ground states, the spectrum of $\frac{3}{2}^-$
molecules is computed. We predict  a strongly bound 15-plet of
$C=-1$ states. A narrow crypto-exotic octet of charm-zero states
is foreseen. In the $C=+1$ sector a sextet of narrow resonances is
formed due to the interaction of D mesons with the baryon
decuplet. A strongly bound triplet of double-charm states is a
consequence of coupled-channel dynamics driven by the D mesons.
\end{abstract}
\end{frontmatter}

\section{Introduction}

In this work we further explore the implications of the
hadrogenesis conjecture \cite{LK01,LK02,LWF02} that baryon
resonances not belonging to the large-$N_c$ ground states are
generated dynamically by coupled-channel dynamics. In recent works
\cite{LK01,LK02,Granada,Copenhagen} it was shown that chiral
symmetry in combination with coupled-channel dynamics
\cite{LK01,LK02} predicts the existence of a wealth of s-wave and
d-wave baryon resonances. The exciting novelty of these studies
lies in the systematics that supports the hadrogenesis conjecture:
there are many resonance states that are naturally recovered
within a coupled-channel computation based on an interaction
unambiguously determined by the chiral properties of QCD.

For the first time it was shown only recently in \cite{Copenhagen}
that d-wave resonances are an immediate consequence of QCD's
chiral SU(3) symmetry. In the early days of hadron physics a few
selected candidates, like the $\Lambda(1405)$, were suggested to
be hadronic molecules, i.e. their masses and partial decay widths
were obtained in terms of a coupled-channel theory with effective
hadronic degrees of freedom rather than quarks and gluons
\cite{Wyld,Dalitz,Ball,Rajasekaran,Wyld2,sw88}. This is quite
analogous to the physics of nuclei, many properties of which are
most economically understood in terms of nucleon degrees of
freedom. The success of the early works
\cite{Wyld,Dalitz,Ball,Rajasekaran,Wyld2,sw88}, that addressed
s-wave states only, was revived by many authors in the last decade
applying a more systematic framework
\cite{ksw95,grnpi,grkl,Oset-prl,Oset-plb,Jido03} but basically
applying an interaction already used decades ago in
\cite{Wyld,Dalitz,Ball,Rajasekaran,Wyld2,sw88}. The close
correspondence of those works is a consequence of the fact that
the leading chiral interaction may be represented by a zero-range
t-channel exchange of vector mesons.

A further important step was taken in
\cite{KL04,LK04-charm,HL04,LK05} where for the first time the
chiral coupled-channel dynamics was claimed to be applicable in
the charm and beauty sectors as well. The crucial observation was
that chiral symmetry predicts unambiguously the s-wave interaction
of the Goldstone bosons with the $\frac{1}{2}^+$ and
$\frac{3}{2}^+$ ground states of non-zero charm quantum numbers.
Coupled-channel dynamics based on the chiral interaction predicted
a zoo of s- and d-wave resonances with charm
\cite{LK04-charm,LK05}.

At present the spectrum of charmed baryon resonances is very
poorly known. It is important to perform detailed computations
that help the discovery of new states. The computation of the
$\frac{1}{2}^-$ spectrum \cite{LK04-charm} was extended recently
for the inclusion of additional channels involving the $\eta'$-,
$\eta_c$- and D-mesons \cite{Hofmann:Lutz:2005}. The study was
based on a force defined by a zero-range t-channel exchange of
vector mesons that is compatible with constraints set by chiral
symmetry. A rich spectrum of $\frac{1}{2}^-$ molecules was
obtained that reproduces known states but predicts plenty of
unknown states, in part with exotic quantum numbers. It is stressed
that those results were not based on assuming a chiral SU(4) symmetry, rather
chiral SU(3) symmetry was realized in terms of universally coupled light
vector mesons. Additional 3-point vertices involving heavy vector mesons
were estimated by a SU(4) assumption. The purpose
of the present study is to work out the analogous generalization
of the $\frac{3}{2}^-$ spectrum. We consider the coupled-channel
interaction of the 16-plet of $0^-$ ground state mesons with the
20-plet of $\frac{3}{2}^+$ baryon ground states.

We recover the spectrum of chiral excitations formed by the
interaction of the Goldstone bosons with baryon ground states
\cite{KL04,Hofmann:Lutz:2005}. Those states decouple to high
accuracy from the coupled-channel dynamics driven by the D mesons.
The latter imply a strongly bound 15-plet of $C=-1$ states and  a
narrow crypto-exotic octet of charm-zero states. In the $C=+1$
sector a sextet of narrow resonances is formed due to the
interaction of D mesons with the baryon decuplet. A strongly bound
triplet of double-charm states is a consequence of coupled-channel
dynamics driven by the D mesons.

\newpage

\section{Coupled-channel interactions}

We study the interaction of the ground-state mesons and baryons
with $J^P= 0^- $ and $\frac{3}{2}^+$ quantum numbers composed out
of u,d,s,c quarks. The pseudoscalar mesons that are considered
can be grouped into multiplet fields $\Phi_{[9]}, \Phi_{[\bar 3]}$
and $\Phi_{[1]}$. We introduce the fields
\begin{eqnarray}
&& \Phi_{[9]}=
\tau \cdot \pi(139) + \alpha^{\dag} \cdot K(494)
+ K^{\dag}(494) \cdot \alpha + \eta(547)\ \lambda_8
+ \sqrt{{\textstyle{2\over 3}}}\,{\one } \,\eta'(958)\,,
\nonumber\\
&& \Phi_{[\bar 3]} = {\textstyle{1\over \sqrt{2}}}\,\alpha^\dagger \cdot D_{\phantom{\mu}}(1867)
- {\textstyle{1\over \sqrt{2}}}\,D^t_{\phantom{\mu}}(1867)\cdot \alpha   +  i\,\tau_2\,D_{\phantom{\mu}}^{(s)}(1969)
\,,
\nonumber\\
&&\Phi_{[1]} = \eta_c(2980)\,,
\label{mesons}
\end{eqnarray}
which are decomposed further into SU(2) multiplets. The
approximate masses in units of MeV as used in the numerical
simulation of this work are recalled in brackets \cite{PDG04}. The
matrices $\tau $ and $\alpha$ are given in terms of the Gell-Mann
SU(3) generators $\lambda_{1,...,8}$ with
\begin{eqnarray}
\tau = (\lambda_1,\lambda_2,\lambda_3), \quad \alpha^{\dagger}=
\frac1{\sqrt2}(\lambda_4+i\lambda_5,\lambda_6+i\lambda_7)\,.
\end{eqnarray}
The baryon states are collected into SU(3) multiplet fields
$B_{[10]}, B_{[6]}, B_{[3]}$ and $B_{[1]}$.  The baryon decuplet
field  is completely symmetric and related to the physical states
by
\begin{eqnarray}
\begin{array}{llll}
B_{[10]}^{111} = \Delta^{++}\,, & B_{[10]}^{113}
=\Sigma^{+}/\sqrt{3}\,, &
B_{[10]}^{133}=\Xi^0/\sqrt{3}\,,  &B_{[10]}^{333}= \Omega^-\,, \\
B_{[10]}^{112} =\Delta^{+}/\sqrt{3}\,, & B_{[10]}^{123}
=\Sigma^{0}/\sqrt{6}\,, &
B_{[10]}^{233}=\Xi^-/\sqrt{3}\,, & \\
B_{[10]}^{122} =\Delta^{0}/\sqrt{3}\,, & B_{[10]}^{223}
=\Sigma^{-}/\sqrt{3}\,, &
& \\
B_{[10]}^{222} =\Delta^{-}\,. & & &
\end{array}
\label{dec-field}
\end{eqnarray}
The sextet, triplet and singlet fields are  decomposed into their
isospin multiplet components
\begin{eqnarray}
&& \sqrt{2}\, B_{[6]} =
{\textstyle{1\over \sqrt{2}}}\,\alpha^{\dag}\cdot \Xi_c(2646)
+ {\textstyle{1\over \sqrt{2}}}\, \Xi_c^t (2646)\cdot \alpha +\Sigma_c(2518)\cdot (i\,\tau \,\tau_2)
\nonumber\\
&& \qquad \;\;\;\;\; + {\textstyle{\sqrt{2}\over 3}}\,
(1-\sqrt{3}\,\lambda_8)\,\Omega_c(2770) \,,
\nonumber\\
&& \sqrt{2}\, B_{[3]} = {\textstyle{1\over
\sqrt{2}}}\,\alpha^\dagger \cdot \Xi_{cc}(3519) -
{\textstyle{1\over \sqrt{2}}}\,\Xi^t_{cc}(3519)\cdot \alpha +
 i\,\tau_2\,\Omega_{cc}(3620) \,,
 \nonumber\\
&& \sqrt{2}\, B_{[1]} = \sqrt{2}\,\Omega_{ccc}(4600) \,,
\label{baryons}
\end{eqnarray}
with their masses taken in this work given in units of MeV. The
$\Omega_c(2770)$, $\Xi_{cc}(3519)$, $\Omega_{cc}(3620)$ and
$\Omega_{ccc}(4600)$ states are not established so far. The mass
of the $\Omega_c(2770)$ is estimated by assuming an equal spacing
rule for the SU(3) breaking pattern of the sextet. The mass of the
$\Xi_{cc}(3159)$ is obtained by the premise that the double-charm
state claimed by the SELEX collaboration
\cite{SELEX,SELEX:Russ,SELEX:Engelfried-a} has $J^P=\frac{3}{2}^+$
quantum numbers. The mass of the $\Omega_{cc}(3620)$ state is
estimated from model calculations
\cite{Rho:Riska:Scoccola:92,Koerner:94} which predict a typical
splitting to the $\Xi_{cc}$ mass of about 100 MeV. Finally the
mass of the $\Omega_{ccc}(4600)$ is guessed from
\cite{Rho:Riska:Scoccola:92,Koerner:94} taking into account the
too large predictions of the $\Omega_{cc}(3620)$ mass.

In a first step we construct the interaction of the mesons and baryon fields introduced in
(\ref{mesons}, \ref{baryons}) with the nonet-field of light vector mesons
\begin{eqnarray}
&& V^{[9]}_\mu = \tau \cdot \rho_\mu(770)
+ \alpha^\dagger \cdot  K_\mu(894) +  K_\mu^\dagger(894) \cdot \alpha
\nonumber\\
&& \qquad +
\Big( {\textstyle{2\over 3}}+ {\textstyle{1\over \sqrt{3}}} \,\lambda_8  \Big)\, \omega_\mu(783)
+\Big(  {\textstyle{\sqrt{2}\over 3}}-\sqrt{{\textstyle{2\over 3}}}\,\lambda_8\Big)\,
\phi_\mu(1020) \,.
\label{light-vectors}
\end{eqnarray}
The multiplet fields are
introduced with
\begin{eqnarray}
&& V_\mu^{[\bar 3]} ={\textstyle{1\over \sqrt{2}}}\,\alpha^\dagger \cdot D_\mu(2008)
- {\textstyle{1\over \sqrt{2}}}\,D^t_\mu(2008)\cdot \alpha   +  i\,\tau_2\,D^{(s)}_{\mu}(2112) \,,
\nonumber\\
&& V_\mu^{[1]} = (J/\Psi)_\mu(3097)\,.
\label{}
\end{eqnarray}
We recall the complete list of SU(3) invariant 3-point vertices
that involve the minimal number of derivatives:
\begin{eqnarray} \label{LM}
\mathcal{L}^{\rm{SU(3)}}_{\rm{int}}
&=& {\textstyle{i\over 4}}\, h_{\bar 3 \bar 3}^9\    {\rm tr}\left(
 (\partial_{\mu}\Phi_{[\bar 3]})\, \Phi^\dagger_{[\bar 3]} V_{[9]}^{\mu} -
  \Phi_{[\bar 3]}\,
(\partial_{\mu}\Phi^\dagger_{[\bar 3]})\,V_{[9]}^{\mu}\right)
\nonumber\\
&+& {\textstyle{i\over 4}}\, h_{\bar 3 \bar 3}^1\    {\rm tr}\left(
(\partial_{\mu}\Phi_{[\bar 3]})\, \Phi^\dagger_{[\bar 3]}-
\Phi_{[\bar 3]}\,(\partial_{\mu}\Phi^\dagger_{[\bar 3]})  \right) \cdot {\rm tr} \left (V_{[9]}^{\mu} \right)
\nonumber\\
&+& {\textstyle{i\over 4}}\, h_{99}^9\    {\rm tr}\left(
(\partial_{\mu}\Phi_{[9]})\, \Phi_{[9]}\, V_{[9]}^{\mu}
-\Phi_{[9]}\,(\partial_{\mu} \Phi_{[9]})\,  V_{[9]}^{\mu} \right)
\nonumber\\
&+& {\textstyle{i\over 4}}\, h_{\bar 3 \bar 3}^0\
{\rm tr}\left(
(\partial_{\mu}\Phi_{[\bar 3]})\, \Phi^\dagger_{[\bar 3]} V_{[1]}^{\mu}
-\Phi_{[\bar 3]}\,(\partial_{\mu}\Phi^\dagger_{[\bar 3]})\, V_{[1]}^{\mu}  \right)
\nonumber\\
&+& {\textstyle{i\over 4}}\, h_{9\bar 3}^{\bar 3}\
{\rm tr}\left(
(\partial_{\mu} \Phi^\dagger_{[\bar 3]})\, \Phi_{[9]}\, V_{[\bar 3]}^{\mu}-
\Phi_{[9]}\,(\partial_{\mu}\Phi_{[\bar 3]})\, V_{[\bar 3]}^{\dagger \mu}  \right)
\nonumber\\
&+& {\textstyle{i\over 4}}\, h_{\bar 39}^{\bar 3}\
{\rm tr}\left(
(\partial_{\mu}\Phi_{[9]})\, \Phi_{[\bar 3]} V_{[\bar 3]}^{\dagger \mu}-
\Phi^\dagger_{[\bar 3]}\,(\partial_{\mu}\Phi_{[9]})\, V_{[\bar 3]}^{\mu}  \right)
\nonumber\\
&+& {\textstyle{i\over 4}}\, h_{\bar 3 0}^{\bar 3}\
{\rm tr}\left(
(\partial_{\mu}\Phi_{[1]})\, \Phi_{[\bar 3]} V_{[\bar 3]}^{\dagger \mu} -
 \Phi^\dagger_{[\bar 3]}\,(\partial_{\mu}\Phi_{[1]})\, V_{[\bar 3]}^{ \mu} \right)
\nonumber\\
&+& {\textstyle{i\over 4}}\, h_{0\bar 3}^{\bar 3}\
{\rm tr}\left(
(\partial_{\mu}\Phi^\dagger_{[\bar 3]}) \,\Phi_{[1]} V_{[\bar 3]}^{\mu}
-\Phi_{[1]}\,(\partial_{\mu}\Phi_{[\bar 3]})\, V_{[\bar 3]}^{\dagger \mu}  \right)
\nonumber\\
&+& {\textstyle{i\over 4}}\, h_{\bar 3 1}^{\bar 3}\
{\rm tr}\left( \Phi^\dagger _{[\bar 3]}\,V_{[\bar 3]}^{\mu}-
\Phi_{[\bar 3]}\,V_{[\bar 3]}^{\dagger \mu}
\right)
{\tr }(\partial_{\mu}\Phi_{[9 ]})\,
\nonumber\\
&+& {\textstyle{i\over 4}}\, h_{1 \bar 3 }^{\bar 3}\ {\rm
tr}\left( (\partial_{\mu} \Phi_{[\bar 3]})\,V_{[\bar
3]}^{\dagger\mu}- (\partial_{\mu} \Phi^\dagger_{[\bar
3]})\,V_{[\bar 3]}^{ \mu} \right) {\tr }(\Phi_{[9 ]})\,.
\label{vector-pseudoscalar}
\end{eqnarray}
It was emphasized in \cite{Hofmann:Lutz:2005} that the terms in
(\ref{vector-pseudoscalar}) are not at odds with the constraints
set by chiral symmetry provided the light vector mesons are
coupled to matter fields via a gauge principle
\cite{Weinberg:68,Bando:85}. The latter requires a correlation of
the coupling constants $h$ in (\ref{vector-pseudoscalar})
\begin{eqnarray}
h_{99}^9 = \frac{(m^{(V)}_{[9]})^2}{2\,g\,f^2}\,, \qquad h_{\bar 3 \bar 3}^{9} = 2\,g \,,
\label{chiral-constraint-mesons}
\end{eqnarray}
with the pion decay constant $f \simeq 92$ MeV. Here the universal
vector coupling strength is $g \simeq 6.6$ and the mass of the
light vector mesons is $m_{[9]}^{(V)}$. A further constraint is
implied by the OZI rule \cite{OZI,Hofmann:Lutz:2005} which claims
\begin{eqnarray}
&& h^1_{\bar 3\bar 3} = -g\,. \label{OZI-constraint}
\end{eqnarray}
The remaining coupling constants are basically  unknown at
present. It was pointed out in \cite{Hofmann:Lutz:2005} that a
SU(4) ansatz is compatible with the chiral and large-$N_c$
constraints (\ref{chiral-constraint-mesons},
\ref{OZI-constraint}), but suggests in addition the relations
\begin{eqnarray}
&&h^0_{\bar 3 \bar 3} = \sqrt{2}\,g\,, \quad
h^{\bar 3}_{9 \bar 3} = 2\,g\,, \quad \;\;\,\,
h^{\bar 3}_{\bar 3 9} = 2\,g\,,
\nonumber\\
&&h^{\bar 3}_{\bar 3 0} = \sqrt{2}\,g\,, \quad
h^{\bar 3}_{0 \bar 3} = \sqrt{2}\,g\,, \quad
h^{\bar 3}_{\bar 3 1} = g\,, \quad \;
h^{\bar 3}_{1 \bar 3} = g \,.
\label{meson-SU4-result}
\end{eqnarray}
A surprising additional consequence of a SU(4) ansatz is the KSFR
relation \cite{KSFRa,KSFRb}
\begin{eqnarray}
\frac{(m^{(V)}_{[9]})^2}{2\,f^2\,g} = g \,.
\label{KSFR-relation}
\end{eqnarray}
Confronting (\ref{meson-SU4-result}) with the scarce empirical
information available suggests a surprisingly small SU(4) breaking
pattern at the level of three-point vertices
\cite{Hofmann:Lutz:2005}. The SU(4) estimates appear to be valid
up to a factor of two. It is reassuring that a similar breaking
pattern is seen also in the weak-decay constants of the
pseudo-scalar mesons \cite{PDG04}.

We continue with the construction of the three-point vertices involving baryon fields.
A complete list of SU(3) invariant terms reads:
\begin{eqnarray}
\mathcal{L}^{\rm{SU(3)}}_{\rm{int}} &=& -{\textstyle{1\over
2}}\,g_{00}^{9}\, {\tr }\Big(\big( \bar B^{[10]}_\nu\,\gamma_\mu
\cdot B^\nu_{[10]}\big) \,V_{[9]}^\mu \Big) -{\textstyle{1\over
2}}\,g_{00}^{1}\, {\tr }\Big( \bar B^{[10]}_\nu\,\gamma_\mu \,
\cdot B^\nu_{[10]} \Big)\, {\tr} \Big( \,V_{[9]}^\mu \Big)
\nonumber\\
&-&{\textstyle{1\over 2}}\,g_{33}^{9}\, {\tr }\Big(\bar
B^{[3]}_\nu\,\gamma_\mu\, B^\nu_{[3]} \,V_{[9]}^\mu\Big)
-{\textstyle{1\over 2}}\,g_{33}^{1}\, {\tr }\Big(\bar
B^{[3]}_\nu\,\gamma_\mu \, B^\nu_{[3]} \Big)\, {\tr} \Big(
\,V_{[9]}^\mu \Big)
\nonumber\\
&-&{\textstyle{1\over 2}}\,g_{66}^{9}\, {\tr }\Big(\bar
B^{[6]}_\nu\,\gamma_\mu \,V_{[9]}^\mu\, B^\nu_{[6]} \Big)
-{\textstyle{1\over 2}}\,g_{66}^{1}\, {\tr }\Big(\bar
B^{[6]}_\nu\,\gamma_\mu \, B^\nu_{[6]} \Big)\, {\tr} \Big(
\,V_{[9]}^\mu \Big)
\nonumber\\
&-& {\textstyle{1\over 2}}\,g_{11}^{1}\, {\tr }\Big(\bar
B_{[1],\nu}\,\gamma_\mu \, B^\nu_{[1]} \Big)\, {\tr} \Big(
\,V_{[9]}^\mu \Big) -{\textstyle{1\over 2}}\,g_{11}^{0}\, {\tr
}\Big(\bar B_{[1],\nu}\,\gamma_\mu \, B^\nu_{[1]}\,V_{[1]}^\mu
\Big)
\nonumber\\
&-& {\textstyle{1\over 2}}\,g_{00}^{0}\, {\tr }\Big( \big( \bar
B^{[10]}_\nu\,\gamma_\mu \,\cdot B^\nu_{[10]}\big) \,
\,V_{[1]}^\mu \Big) -{\textstyle{1\over 2}}\,g_{33}^{0}\, {\tr
}\Big(\bar B^{[3]}_\nu\,\gamma_\mu  B^\nu_{[3]} \,V_{[1]}^\mu
\Big)
\nonumber\\
&-& {\textstyle{1\over 2}}\,g_{66}^{0}\, {\tr }\Big( \bar
B^{[6]}_\nu\,\gamma_\mu \,B^\nu_{[6]} \,V_{[1]}^\mu \Big)
\nonumber\\
&-&{\textstyle{1\over 2}}\ g_{63}^{\bar 3}\ {\rm{tr}}\ \Big( \bar
B^{[6]}_\nu\,\gamma_{\mu} \, {\rm{tr}} \big( B^\nu_{[3]}\,V_{[\bar
3]}^{\dagger\mu} \big) -2\, \bar
B^{[6]}_\nu\,\gamma_{\mu}\,B^\nu_{[3]}\,V_{[\bar 3]}^{\dagger\mu}
\nonumber\\
& & \qquad \;\;\;\;+ \bar B^{[3]}_\nu\,\gamma_{\mu}\,V_{[\bar
3]}^{\phantom{\dagger}\mu} \,{\rm{tr}}\, \big( B^\nu_{[6]} \big)
-2\,\bar B^{[3]}_\nu\,\gamma_{\mu}\, B^\nu_{[6]}\,V_{[\bar
3]}^{\phantom{\dagger}\mu} \Big)
\nonumber\\
&-& {\textstyle{1\over 2}}\,g_{31}^{\bar 3}\, {\tr }\Big(\bar
B_{[1],\nu}\,\gamma_\mu \, B^\nu_{[3]} \, V^{\dagger\, \mu}_{[3]}
+\bar B^{[3]}_\nu\,\gamma_\mu \, B^\nu_{[1]} \,V_{[\bar 3]}^\mu
\Big)
\nonumber\\
&-& {\textstyle{1\over 2}}\,g_{06}^{\bar 3}\, {\tr }\Big( \big(
\bar B^{[10]}_\nu\,\gamma_\mu \, \cdot B^\nu_{[6]} \big) \,
V^{\dagger\, \mu}_{[3]} -\big( \bar B^{[6]}_\nu\,\gamma_\mu \,
\cdot B^\nu_{[10]} \big) \,V_{[\bar 3]}^\mu \Big)\,,
 \label{vector-baryons}
\end{eqnarray}
where we apply the notation
\begin{eqnarray}
&&\big[\bar B_{[10]} \cdot B_{[6]}\big]^{ab} = \epsilon^{ab
k}\,\bar B_{[10], ijk} \cdot B_{[6]}^{ij} \,, \qquad \big[\bar
B_{[6]} \cdot B_{[10]}\big]_{ab} = \epsilon_{ab k}\,\bar B_{[6],
ij} \cdot B_{[10]}^{ijk} \,, \nonumber\\
&& \big[\bar B_{[10]} \cdot B_{[10]}\big]^{a}_{b} = \bar
B_{[10],ijb}\, B_{[10]}^{ija} \,.
 \label{}
\end{eqnarray}
Within the hidden local symmetry model \cite{Bando:85} chiral
symmetry is recovered with
\begin{eqnarray}
g_{00}^9= 3\,g \,, \qquad g_{66}^9=-g_{33}^9=2\,g \,.
\label{chiral-constraint-baryons}
\end{eqnarray}
It is acknowledged that chiral symmetry does not constrain the
coupling constants in (\ref{vector-baryons}) involving the SU(3)
singlet part of the fields. The latter can, however, be
constrained  by a large-$N_c$ operator analysis \cite{DJM}. At
leading order in the $1/N_c$ expansion the OZI rule \cite{OZI} is
predicted.  As a consequence the estimates
\begin{eqnarray}
g_{00}^1=g_{11}^1= g_{66}^1=0 \,,\qquad g_{33}^1=g \,,
\label{OZI-constraint-baryons}
\end{eqnarray}
follow.

It is instructive to construct also the
SU(4) symmetric generalization of the interaction (\ref{vector-baryons}). The baryons
form a 20-plet in SU(4). Its field is represented by a tensor $B^{ijk}$, which is completely
symmetric. The indices $i,j,k$ run from one to four, where one can read off the
quark content of a baryon state by the identifications $1 \leftrightarrow u,
2 \leftrightarrow d, 3 \leftrightarrow s, 4 \leftrightarrow c$.
We write:
\begin{eqnarray}
&&{\mathcal L}^{\rm{SU(4)}}_{\rm{int}} = -{\textstyle{3\over
4}}\,g \sum_{i,j,k,l=1}^4\, \bar B^{[20]}_{ijk} \,\gamma^\mu\,
V_{\mu,\,l}^{[16],k}\,B^{ijl}_{[20]}\,,
\label{baryons-SU4} \\
&& \renewcommand{\arraystretch}{1.3}
\begin{array}{llll}
 B_{[20]}^{111}=\Delta^{++} \,, &    B_{[20]}^{112}=\frac1{\sqrt3}\,\Delta^{+}\,, &  B_{[20]}^{122}=\frac1{\sqrt3}\, \Delta^0\,, &   B_{[29]}^{222}=\Delta^-\,,\\*
 B_{[20]}^{113}=\frac1{\sqrt3}\Sigma^+ \,, &  B_{[20]}^{123}=\frac1{\sqrt6}\Sigma^0 \,, & B_{[20]}^{223}=\frac1{\sqrt3}\Sigma^-\,, &    B_{[20]}^{133}=\frac1{\sqrt3}\Xi^0 \,,\\*
 B_{[20]}^{233}=\frac1{\sqrt3}\Xi^-\,, &    B_{[20]}^{333}=\Omega^- \,, &     B_{[20]}^{114}=\frac1{\sqrt3}\Sigma_c^{++}\,, &  B_{[20]}^{124}=\frac1{\sqrt6}\Sigma_c^+\,,\\*
 B_{[20]}^{224}=\frac1{\sqrt3}\Sigma_c^0 \,, &  B_{[20]}^{134}=\frac1{\sqrt6}\Xi_c^+ \,, &  B_{[20]}^{234}=\frac1{\sqrt6} \Xi_c^0 \,, &  B_{[20]}^{334}=\frac1{\sqrt3}\Omega_c^0\,,\\*
 B_{[20]}^{144}=\frac1{\sqrt3}\Xi_{cc}^{++} \,, &  B_{[20]}^{244}=\frac1{\sqrt3}\Xi_{cc}^+ \,, &  B_{[20]}^{344}=\Omega_{cc}^+ \,, &   B_{[20]}^{444}=\Omega_{ccc}^{++} \,,
\end{array}
\end{eqnarray}
where we use the vector field $V_{[16]}^\mu$ as introduced in
\cite{Hofmann:Lutz:2005}.
 It is pointed out that the relations
(\ref{chiral-constraint-baryons}, \ref{OZI-constraint-baryons})
follow from the form of the interaction (\ref{baryons-SU4}). In
addition the SU(4) symmetric vertex (\ref{baryons-SU4}) implies
\begin{eqnarray}
&&g_{06}^{\bar 3}=\sqrt{\frac32}\,g \,,\qquad g_{63}^{\bar 3}=
2\,g \,,\qquad
g_{31}^{\bar 3}= -\sqrt3\,g \,, \nonumber\\
&& g_{00}^0=0\,,\qquad g_{66}^0=\sqrt2 g\,, \qquad
g_{33}^0=2\,\sqrt2 g\,, \qquad g_{11}^0=3\,\sqrt2g\,.
\label{baryon-SU4-result}
\end{eqnarray}
Unfortunately there appears to be no way at present to check on
the usefulness of the result (\ref{baryon-SU4-result}). Eventually
simulations of QCD on a lattice may shed some light on this issue.
The precise values of the coupling constants
(\ref{meson-SU4-result}, \ref{baryon-SU4-result}) will not affect
the major results of this work. This holds as long as those
coupling constants range in the region suggested by
(\ref{meson-SU4-result}, \ref{baryon-SU4-result}) within a factor
three.

\newpage

\section{Coupled-channel scattering}

\begin{table}[b]\rescale
\setlength{\tabcolsep}{1.4mm}
\setlength{\arraycolsep}{1.65mm}
\renewcommand{\arraystretch}{1.2}
\begin{center}
\tabcolsep=2.4mm
\begin{tabular}{|@{}c@{}c@{}c@{}c@{}c@{}c@{}c@{}c@{}c@{}c@{}c@{}c@{}|}
\hline \hline
\multicolumn{3}{|c|}{\makebox[3cm][c]{$(1,0,-1)$}} &
\multicolumn{3}{c|}{\makebox[3cm][c]{$(2,0,-1)$}} &
\multicolumn{3}{c|}{\makebox[3cm][c]{$(\frac12,-1,-1)$}} &
\multicolumn{3}{c|}{\makebox[3cm][c]{$(\frac32,-1,-1)$}}\\
\hline \multicolumn{3}{|c|}{$(\sqrt{\frac34} \bar D^t
\,i\,\sigma_2 \vec T^\dag \Delta)$} &
  \multicolumn{3}{c|}{\makebox[3.1cm][c]{$ (\sqrt{\frac18} \bar D^t \,i\,\sigma_2 (\sigma_i\,T_j^\dag\!+\sigma_j\,T_i^\dag)\Delta ) $}}&
\multicolumn{3}{c|}{$ (\frac1{\sqrt3}\Sigma \cdot \sigma \bar{D}) $} &
\multicolumn{3}{c|}{$ \left(\begin{array}{c} (\bar D_s \Delta) \\ (\Sigma \cdot T \,\bar D)\end{array}\right) $}\\
\hline \hline
\multicolumn{3}{|c|}{\makebox[3cm][c]{$(0,-2,-1)$}} &
\multicolumn{3}{c|}{\makebox[3cm][c]{$(1,-2,-1)$}} &
\multicolumn{3}{c|}{\makebox[3cm][c]{$(\frac12,-3,-1)$}} &
\multicolumn{3}{c|}{\makebox[3cm][c]{$(0,-4,-1)$}}\\
\hline \multicolumn{3}{|c|}{$ (\frac{i}{\sqrt2} \bar D^t
\,\sigma_2 \,\Xi) $} &
\multicolumn{3}{c|}{$\left(\begin{array}{c}  (\bar{D}_s \,\vec{\Sigma}) \\
(\frac{i}{\sqrt2} \bar{D}^t \,\sigma_2 \vec{\sigma}\, \Xi)
\end{array}\right) $} &
\multicolumn{3}{c|}{$\left(\begin{array}{c} (\bar{D}_s \,\Xi) \\
(\bar D \, \Omega) \end{array}\right) $} &
\multicolumn{3}{c|}{$(\bar D_s \Omega)$} \\ \hline \hline
\end{tabular}
\caption{Coupled-channel states with isospin ($I$), strangeness
($S$) and charm ($C=-1$). } \label{tab:states-1}
\end{center}
\end{table}

\begin{table}[t]\rescale
\setlength{\tabcolsep}{1.4mm} \setlength{\arraycolsep}{1.65mm}
\renewcommand{\arraystretch}{1.2}
\begin{center}
\tabcolsep=1.5mm
\begin{tabular}{|@{}c@{}c@{}c@{}c@{}c@{}c@{}c@{}c@{}c@{}c@{}c@{}c@{}|}
\hline \hline \multicolumn{3}{|c|}{\makebox[3cm][c]{$(1,1,0)$}} &
\multicolumn{3}{c|}{\makebox[3cm][c]{$(2,1,0)$}} &
\multicolumn{3}{c|}{\makebox[3cm][c]{$(\frac12,0,0)$}} &
\multicolumn{3}{c|}{\makebox[3cm][c]{$(\frac32,0,0)$}}\\
\hline 
\multicolumn{3}{|c|}{$ (\sqrt{\frac34} K^t i\,\sigma_2 \,\vec
T^\dag \Delta) $} & \multicolumn{3}{c|}{\makebox[3.1cm][c]{$
(\frac{i}{\sqrt8}K^t\,\sigma_2(\sigma_i\,T_j^\dag+\sigma_j\,T_i^\dag)\Delta)
$}} &
\multicolumn{3}{c|}{$\left(\begin{array}{c} (\frac{1}{\sqrt2} \pi \cdot T^\dag \Delta) \\
(\frac{1}{\sqrt3} \,\Sigma \cdot \sigma \,K) \\
(\frac{1}{\sqrt3} \,\Sigma_c \cdot \sigma \,\bar{D})
\end{array}\right) $} &
\multicolumn{3}{c|}{$ \left(\begin{array}{c} ({\textstyle{3\over
\sqrt{15}}}\,T_i \,(\pi
\cdot \sigma ) \,T_i^\dagger\, \Delta ) \\
(\eta\,\Delta) \\
(\Sigma \cdot T \,K) \\
(\eta' \Delta) \\
(\eta_c \Delta) \\
(\Sigma_c \cdot T \,\bar{D}) \end{array}\right) $}\\
\hline \hline \multicolumn{4}{|c|}{$(\frac52,0,0)$} &
\multicolumn{4}{c|}{$(0,-1,0)$} &
\multicolumn{4}{c|}{$(1,-1,0)$}\\
\hline \multicolumn{4}{|c|}{$
\begin{array}{c}
\big( (\frac{1}{2}\,(\pi_i\,T^\dagger_j + \pi_j\,T^\dagger_i )- \frac{1}{3}\,\delta_{ij}\, (T^\dagger \cdot \pi) \\
- \frac{1}{10}\, (\sigma_i\,T^\dagger_j+\sigma_j\,T^\dagger_i
)\,T_l\,(\sigma \cdot \pi)\,T^\dagger_l ) \,\Delta \big)
\end{array} $} &
\multicolumn{4}{|c|}{$ \left( \begin{array}{c} (\frac1{\sqrt3}\pi\cdot\Sigma) \\
(\frac{i}{\sqrt2} K^t \sigma_2 \,\Xi) \\
(\frac{i}{\sqrt2} \bar D^t \sigma_2 \,\Xi_c) \end{array}\right) $}
&
\multicolumn{4}{c|}{$\left(\begin{array}{c} (\frac{i}{\sqrt2} \vec{\Sigma} \times \vec{\pi}) \\
(\sqrt{\frac34}\bar K \,\vec T^\dag \Delta ) \\
(\eta \,\vec{\Sigma}) \\
(\frac{i}{\sqrt2} K^t\, \sigma_2 \vec{\sigma}\, \Xi) \\
(\eta' \vec{\Sigma}) \\
(\eta_c \vec{\Sigma}) \\
(\bar{D}_s \vec{\Sigma}_c) \\
(\frac{i}{\sqrt2} \bar{D}^t\, \sigma_2 \vec{\sigma}\, \Xi_c)
\end{array}\right) $} \\
\hline \hline
\multicolumn{4}{|c|}{\makebox[4.15cm][c]{$(2,-1,0)$}} &
\multicolumn{4}{c|}{\makebox[4.15cm][c]{$(\frac12,-2,0)$}} &
\multicolumn{4}{c|}{\makebox[4.15cm][c]{$(\frac32,-2,0)$}}\\
\hline
\multicolumn{4}{|c|}{\makebox[3.5cm][c]{$\left(\begin{array}{c}
(\frac12\,(\pi_i\Sigma_j+\pi_j\Sigma_i)-\frac{\delta_{ij}}{3}\pi\cdot\Sigma) \\
(\frac1{\sqrt8}\bar K\, (\sigma_iT_j^\dag+\sigma_jT_i^\dag)\Delta)
\end{array}\right) $}} &
\multicolumn{4}{|c|}{$\rule{0pt}{4ex} \left(\begin{array}{c}
(\frac1{\sqrt3}\pi\cdot\sigma\, \Xi) \\
(\frac{i}{\sqrt3}\Sigma\cdot \sigma\,\sigma_2\bar K) \\
(\eta\, \Xi) \\
(K\Omega) \\
(\eta' \Xi) \\
(\eta_c \Xi) \\
(\bar D_s \Xi_c) \\
(\bar D \Omega_c)
\end{array}\right) $} &
\multicolumn{4}{|c|}{$\left( \begin{array}{c}
(\pi\cdot T\, \Xi) \\
(\Sigma\cdot T i\,\sigma_2\bar K^t)
\end{array}\right) $}\\ \hline \hline
\multicolumn{4}{|c|}{\makebox[4cm][c]{$(0,-3,0)$}} &
\multicolumn{4}{c|}{\makebox[4cm][c]{$(1,-3,0)$}} &
\multicolumn{4}{c|}{\makebox[4cm][c]{$(\frac12,-4,0)$}}\\
\hline \multicolumn{4}{|c|}{$\left(\begin{array}{c}
(\frac1{\sqrt2}\bar K\, \Xi) \\
(\eta\, \Omega) \\
(\eta' \Omega) \\
(\eta_c \Omega) \\
(\bar D_s\Omega_c)
\end{array}\right) $} &
\multicolumn{4}{c|}{$ \left(\begin{array}{c}
(\vec \pi\, \Omega)\\
(\frac1{\sqrt2}\bar K\vec\sigma\, \Xi)
\end{array}\right) $}&
\multicolumn{4}{c|}{$ (\Omega i\,\sigma_2 \bar K^t) $}\\
\hline \hline
\end{tabular}
\caption{Coupled-channel states with isospin ($I$), strangeness
($S$) and charm ($C=0$). } \label{tab:states0}
\end{center}
\end{table}

\begin{table}[t]\rescale
\setlength{\tabcolsep}{1.4mm}
\setlength{\arraycolsep}{1.65mm}
\renewcommand{\arraystretch}{1.2}
\begin{center}
\tabcolsep=0.8mm
\begin{tabular}{|@{}c@{}c@{}c@{}c@{}c@{}c@{}c@{}c@{}c@{}c@{}c@{}c@{}|}
\hline \hline
\multicolumn{3}{|c|}{\makebox[3cm][c]{$(\frac12,1,1)$}} &
\multicolumn{3}{c|}{\makebox[3cm][c]{$(\frac32,1,1)$}} &
\multicolumn{3}{c|}{\makebox[3cm][c]{$(0,0,1)$}} &
\multicolumn{3}{c|}{\makebox[3cm][c]{$(1,0,1)$}}\\
\hline
\multicolumn{3}{|c|}{$(\frac1{\sqrt3}\Sigma_c\cdot\sigma K) $}&
\multicolumn{3}{c|}{$\left(\begin{array}{c}
(\Sigma_c \cdot T\, K)\\
(D_s\Delta)
\end{array}\right) $} &
\multicolumn{3}{c|}{$\left(\begin{array}{c}
(\frac1{\sqrt3}\pi \cdot \Sigma_c) \\
(\frac{i}{\sqrt2}K^t \sigma_2\Xi_c) \\
(\frac{i}{\sqrt2}\bar D^t\sigma_2\Xi_{cc} )
\end{array}\right) $} &
\multicolumn{3}{c|}{$\left(\begin{array}{c}
(\frac{-i}{\sqrt2}\vec{\pi}\times\vec{\Sigma}_c) \\
(\eta\, \vec{\Sigma}_c) \\
(\sqrt{\frac34} D\, \vec T^\dag \Delta) \\
(\frac{i}{\sqrt2}K^t \sigma_2 \vec{\sigma}\, \Xi_c) \\
(D_s\vec{\Sigma}) \\
(\eta' \vec{\Sigma}_c) \\
(\frac{i}{\sqrt2} \bar D^t\sigma_2 \vec{\sigma} \Xi_{cc}) \\
(\eta_c \vec{\Sigma}_c)
\end{array}\right) $} \\
\hline
\hline
\multicolumn{3}{|c|}{\makebox[3cm][c]{$(2,0,1)$}} &
\multicolumn{3}{c|}{\makebox[3cm][c]{$(\frac12,-1,1)$}} &
\multicolumn{3}{c|}{\makebox[3cm][c]{$(\frac32,-1,1)$}} &
\multicolumn{3}{c|}{\makebox[3cm][c]{$(0,-2,1)$}}\\
\hline
\multicolumn{3}{|c|}{$\left(\begin{array}{c} \!\!\!\!\!(\frac12(\pi_i\Sigma_{c,j}\!+\pi_j\Sigma_{c,i})- \!\frac{\delta_{ij}}{3}\pi\cdot\Sigma_c)\!\!\!\!\!\\
(\frac1{\sqrt8} D (\sigma_iT_j^\dag+\sigma_jT_i^\dag)\Delta)
\end{array}\right) $}&
\multicolumn{3}{|c|}{$\left(\begin{array}{c}
(\frac1{\sqrt3}\pi\cdot\sigma\Xi_c)\\
(\frac{i}{\sqrt3}\Sigma_c\cdot\sigma \sigma_2 \bar K^t)\\
(\eta\,\Xi_c) \\
(\frac{i}{\sqrt3}\Sigma\cdot\sigma \sigma_2 D^t)\\
(\Omega_c K) \\
(D_s \Xi) \\
(\eta' \Xi_c) \\
(\Omega_{cc} \bar D) \\
(\bar D_s \Xi_{cc}) \\
(\eta_c\Xi_c)
\end{array}\right) $}&
\multicolumn{3}{|c|}{$\left(\begin{array}{c}
(\pi \cdot T\, \Xi_c) \\
(\Sigma_c\cdot T\,i\,\sigma_2\bar K^t) \\
(\Sigma\cdot T \,i\,\sigma_2 D^t)
\end{array}\right) $}&
\multicolumn{3}{|c|}{$\left(\begin{array}{c}
(\frac1{\sqrt2} \bar K\, \Xi_c) \\
(\eta\, \Omega_c) \\
(\frac1{\sqrt2} D\, \Xi) \\
(D_s \Omega) \\
(\eta' \Omega_c) \\
(\bar D_s \Omega_{cc}) \\
(\eta_c \Omega_c)
\end{array}\right) $} \\
\hline
\hline
\multicolumn{6}{|c|}{\makebox[6cm][c]{$(1,-2,1)$}} &
\multicolumn{6}{c|}{\makebox[6cm][c]{$(\frac12,-3,1)$}} \\
\hline
\multicolumn{6}{|c|}{$\left(\begin{array}{c}
(\vec \pi\, \Omega_c)\\
(\frac1{\sqrt2} \bar K\, \vec\sigma\,\Xi_c) \\
(\frac1{\sqrt2} D\, \vec\sigma\, \Xi)
\end{array}\right) $}&
\multicolumn{6}{c|}{$\left(\begin{array}{c}
(i\,\sigma_2\bar K^t \Omega_c) \\
(i\,\sigma_2 D^t \Omega)
\end{array}\right) $}\\
\hline
\hline
\end{tabular}
\caption{Coupled-channel states with isospin ($I$), strangeness
($S$) and charm ($C=1$). } \label{tab:states1}
\end{center}
\end{table}

\begin{table}[t]\rescale
\setlength{\tabcolsep}{1.4mm}
\setlength{\arraycolsep}{1.65mm}
\renewcommand{\arraystretch}{1.2}
\begin{center}
\tabcolsep=2mm
\begin{tabular}{|@{}c@{}c@{}c@{}c@{}c@{}c@{}c@{}c@{}c@{}c@{}c@{}c@{}|}
\hline \hline
\multicolumn{3}{|c|}{\makebox[3cm][c]{$(0,1,2)$}} &
\multicolumn{3}{c|}{\makebox[3cm][c]{$(1,1,2)$}} &
\multicolumn{3}{c|}{\makebox[3cm][c]{$(\frac12,0,2)$}} &
\multicolumn{3}{c|}{\makebox[3cm][c]{$(\frac32,0,2)$}} \\
\hline
\multicolumn{3}{|c|}{$\left(\begin{array}{c} \frac{i}{\sqrt2}K^t\sigma_2 \Xi_{cc}  \end{array}\right) $}&
\multicolumn{3}{c|}{$\left(\begin{array}{c} (\frac{i}{\sqrt2}K^t\sigma_2\vec{\sigma}\, \Xi_{cc}) \\ (D_s\Sigma_c)  \end{array}\right) $}&
\multicolumn{3}{c|}{$\left(\begin{array}{c}
(\frac1{\sqrt3}\pi\cdot\sigma\Xi_{cc}) \\
(\eta\, \Xi_{cc}) \\
(\Omega_{cc} K) \\
(\frac{i}{\sqrt3}\Sigma_c\cdot\sigma \sigma_2 D^t) \\
(\eta' \Xi_{cc}) \\
(D_s \Xi_c) \\
(\bar D\, \Omega_{ccc}) \\
(\eta_c \Xi_{cc})
\end{array}\right) $}&
\multicolumn{3}{c|}{$\left(\begin{array}{c} (\pi\cdot T\,\Xi_{cc}) \\ (\Sigma_c\cdot T \,i\,\sigma_2 D^t)  \end{array}\right) $}\\
\hline
\hline
\multicolumn{4}{|c|}{\makebox[4cm][c]{$(0,-1,2)$}} &
\multicolumn{4}{c|}{\makebox[4cm][c]{$(1,-1,2)$}} &
\multicolumn{4}{c|}{\makebox[4cm][c]{$(\frac12,-2,2)$}} \\
\hline
\multicolumn{4}{|c|}{$\left(\begin{array}{c}
(\frac1{\sqrt2}\bar K\, \Xi_{cc})\\
(\eta\, \Omega_{cc})\\
(\frac1{\sqrt2}D\,\Xi_c)\\
(\eta'\Omega_{cc})\\
(D_s\Omega_c)\\
(\bar D_s \Omega_{ccc})\\
(\eta_c\Omega_{cc})
\end{array}\right) $}&
\multicolumn{4}{|c|}{$\left(\begin{array}{c}
(\vec\pi\, \Omega_{cc})\\
(\frac1{\sqrt2}\bar K\, \vec\sigma\,\Xi_{cc})\\
(\frac1{\sqrt2}D\,\vec\sigma\,\Xi_c)
\end{array}\right) $}&
\multicolumn{4}{|c|}{$\left(\begin{array}{c}
(i\,\sigma_2\bar K^t \Omega_{cc}) \\
(i\,\sigma_2 D^t \Omega_c)
\end{array}\right) $}\\
\hline
\hline
\multicolumn{3}{|c|}{\makebox[3cm][c]{$(\frac12,1,3)$}} &
\multicolumn{3}{c|}{\makebox[3cm][c]{$(0,0,3)$}} &
\multicolumn{3}{c|}{\makebox[3cm][c]{$(1,0,3)$}} &
\multicolumn{3}{c|}{\makebox[3cm][c]{$(\frac12,-1,3)$}} \\
\hline
\multicolumn{3}{|c|}{$\left(\begin{array}{c} (K\Omega_{ccc}) \\ (D_s\Xi_{cc})  \end{array}\right) $}&
\multicolumn{3}{c|}{$\left(\begin{array}{c}
(\eta\, \Omega_{ccc})\\
(\frac1{\sqrt2}D\,\Xi_{cc})\\
(\eta' \Omega_{ccc})\\
(D_s\Omega_{cc})\\
(\eta_c\Omega_{ccc})
\end{array}\right) $}&
\multicolumn{3}{c|}{$\left(\begin{array}{c} (\vec\pi\, \Omega_{ccc}) \\ (\frac1{\sqrt2}D\,\Xi_{cc})  \end{array}\right) $}&
\multicolumn{3}{c|}{$\left(\begin{array}{c} (\Omega_{ccc} i\,\sigma_2 \bar K^t) \\ (\Omega_{cc}i\,\sigma_2 D^t)  \end{array}\right) $}\\
\hline
\hline
\multicolumn{6}{|c|}{\makebox[6cm][c]{$(0,1,4)$}} &
\multicolumn{6}{c|}{\makebox[6cm][c]{$(\frac12,0,4)$}} \\
\hline
\multicolumn{6}{|c|}{$( D_s\Omega_{ccc}) $}&
\multicolumn{6}{c|}{$( i\,\sigma_2D^t\Omega_{ccc} ) $}\\
\hline
\hline
\end{tabular}
\caption{Coupled-channel states with isospin ($I$), strangeness
($S$) and charm ($C=2,3,4$).} \label{tab:states2-4}
\end{center}
\end{table}

We consider the s-wave scattering of the pseudo-scalar mesons
fields (\ref{mesons}) off the baryon fields (\ref{dec-field}, \ref{baryons}). The
scattering process is described by the amplitudes that follow as
solutions of the Bethe-Salpeter equation,
\begin{eqnarray}
T_{\mu \nu}(\bar k ,k ;w ) &=& K_{\mu \nu}(\bar k ,k ;w )
+\int\!\! \frac{d^4l}{(2\pi)4}\,K_{\mu \alpha}(\bar k , l;w )\,
G_{\alpha \beta}(l;w)\,T_{\beta \nu}(l,k;w )\;,
\nonumber\\
G_{\mu \nu }(l;w)&=&-i\,D(\half\,w-l)\,S_{\mu \nu}( \half\,w+l)\,,
\label{BS-coupled}
\end{eqnarray}
where we suppress the coupled-channel structure for simplicity.
The meson and decuplet propagators,  $D(q)$ and $S_{\mu \nu}(p)$,
are used in the notation of \cite{LWF02}.  We introduced
convenient kinematics:
\begin{eqnarray}
w = p+q = \bar p+\bar q\,, \quad k= \half\,(p-q)\,,\quad \bar k
=\half\,(\bar p-\bar q)\,, \label{def-moment}
\end{eqnarray}
where $q,\,p,\, \bar q, \,\bar p$ are the initial and final meson
and baryon 4-momenta.

The scattering kernel is approximated by the t-channel vector
meson exchange force defined by (\ref{vector-pseudoscalar},
\ref{vector-baryons}), where we apply the formalism developed in
\cite{LK02,LK04-axial}. The scattering kernel has the form
\begin{eqnarray}
K_{\mu \nu}^{(I,S,C)}(\bar k,k;w) = \frac{1}{4}\,\sum_{V \in
[16]}\, \frac{C^{(I,S,C)}_V}{t-m^2_V} \,\Big( \frac{\barqslash
+\qslash }{2} - (\bar
q^2-q^2)\,\frac{\barqslash-\qslash}{2\,m_V^2}\Big)\,g_{\mu \nu}\,,
\label{def-K}
\end{eqnarray}
with  $t=(\bar q-q)^2$.

In (\ref{def-K}) the scattering is projected onto sectors with
conserved isospin (I), strangeness (S) and charm (C) quantum
numbers. The latter are introduced with respect to the states
collected in Tabs. \ref{tab:states-1}-\ref{tab:states2-4}, where
we use the notation of \cite{LK02,LWF02}. The Pauli matrices
$\sigma_i$ act on the isospin doublet fields, like $\Xi$ or $K$.
The $2\times 4$ transition matrices $T$ are normalized according
to
\begin{eqnarray}
\vec T \cdot \vec T^\dagger =1\,, \qquad T^\dag_i
T_j=\delta_{ij}-{\textstyle{1\over 3}}\,\sigma_i\,\sigma_j \,.
\label{def-etc}
\end{eqnarray}
The coupled-channel structure of the matrices $C_{V,ab}^{(I,S,C)}$
is detailed in the Appendix. Only non-vanishing elements are
displayed. Owing to the 'chiral' identifications
(\ref{chiral-constraint-mesons}, \ref{chiral-constraint-baryons})
and the KSFR relation (\ref{KSFR-relation}) we reproduce the
coupled-channel structure of the Weinberg-Tomozawa interaction
identically. Summing over the light vector meson states
\begin{eqnarray}
\sum_{V\in [9]}\,C^{(I,S,C)}_V =4\,g^2\,C_{WT}^{(I,S,C)} \,,
\label{WT-repr}
\end{eqnarray}
we reproduce the matrices $C_{WT}$ as given previously in
\cite{Copenhagen}. The first term of the interaction kernel
matches corresponding expressions predicted by the leading order
chiral Lagrangian if we put $t=0$ in (\ref{def-K}) and use the
common value for the vector-meson masses suggested by the KSFR
relation (\ref{KSFR-relation}). The second term in (\ref{def-K})
is formally of chiral order $Q^3$ for channels involving Goldstone
bosons. Numerically it is a minor correction but nevertheless it
is kept in the computation.

Given (\ref{meson-SU4-result}, \ref{baryon-SU4-result}) one may
decompose the interaction into SU(4) invariant tensors:
\begin{eqnarray}
&&\frac{1}{4\,g^2}\sum_{V\in[16]}\,C^{(I,S,C)}_V = 7\,C_{[20]}+
4\,C_{[20_s]} -3\,C_{[120]}+C_{[140]} \,,
\nonumber\\ \nonumber\\
&& \qquad \qquad 15 \otimes 20 = 20 \oplus 20_s \oplus
\overline{120}  \oplus 140 \,. \label{SU4-decomposition}
\end{eqnarray}
The normalization of the matrices $C_{[...]}$ is such that their
weight factors in (\ref{SU4-decomposition}) give the eigenvalues
of $\sum_V \,C_V /(4\,g^2)$ . Strongest attraction is foreseen in
the $20$-plets, repulsion in the 120-plet. It is interesting to
observe that (\ref{SU4-decomposition}) predicts weak attraction in
the $140$-plet. This is an exotic multiplet. To digest this
abstract group theory we further decompose the SU(4) multiplets
into the more familiar SU(3) multiplets
\begin{eqnarray}
&&[20 ]^{\rm{SU(4)}} = \;\,[8]^{\rm{SU(3)}}_{C=0\,} \oplus
[6]^{\rm{SU(3)}}_{C=1} \oplus [\overline{3}]^{\rm{SU(3)}}_{C=1}
\oplus [3]^{\rm{SU(3)}}_{C=2}  \,,
\nonumber\\
&&[20 ]_{s}^{\rm{SU(4)}} = [10]^{\rm{SU(3)}}_{C=0} \oplus
[6]^{\rm{SU(3)}}_{C=1} \oplus [3]^{\rm{SU(3)}}_{C=2} \oplus
[1]^{\rm{SU(3)}}_{C=3} \,,
\nonumber\\
&&[120 ]^{\rm{SU(4)}} = \;\,[15]^{\rm{SU(3)}}_{C=-1} \oplus
[10]^{\rm{SU(3)}}_{C=0} \oplus [35]^{\rm{SU(3)}}_{C=0} \oplus
[6]^{\rm{SU(3)}}_{C=1} \oplus [24]^{\rm{SU(3)}}_{C=1}\oplus
\nonumber\\
&& \qquad \qquad \quad  [3]^{\rm{SU(3)}}_{C=2}\oplus
[15]^{\rm{SU(3)}}_{C=2} \oplus [1]^{\rm{SU(3)}}_{C=3} \oplus
[8]^{\rm{SU(3)}}_{C=3} \oplus [\bar 3]^{\rm{SU(3)}}_{C=4}\,,
\nonumber\\
&&[140 ]^{\rm{SU(4)}} = \;\,[15]^{\rm{SU(3)}}_{C=-1} \oplus
[8]^{\rm{SU(3)}}_{C=0}\oplus [10]^{\rm{SU(3)}}_{C=0} \oplus
[27]^{\rm{SU(3)}}_{C=0} \oplus [\bar 3]^{\rm{SU(3)}}_{C=1}\oplus
\nonumber\\
&& \qquad \qquad \quad  [6]^{\rm{SU(3)}}_{C=1}\oplus
[\overline{15}]^{\rm{SU(3)}}_{C=1} \oplus
[24]^{\rm{SU(3)}}_{C=1}\oplus[3]^{\rm{SU(3)}}_{C=2} \oplus [\bar
6]^{\rm{SU(3)}}_{C=2}\oplus
\nonumber\\
&& \qquad \qquad \quad [15]^{\rm{SU(3)}}_{C=2} \oplus
[8]^{\rm{SU(3)}}_{C=3} \,.\label{36-decomp}
\end{eqnarray}
It should be stressed that the decomposition
(\ref{SU4-decomposition}) is useful only to perform some
consistency checks of the computation. A SU(4) decomposition
ignores the important physics whether possible attraction is
provided by the t-channel exchange of the light or heavy vector
mesons.

Following \cite{LK02,LWF02,Copenhagen} an effective interaction
kernel is introduced that can be decomposed into a set of
covariant projectors with well defined total angular momentum,
$J$, and parity, $P$,
\begin{eqnarray}
&& V_{\mu \nu}(\bar k ,k ;w )  =
\sum_{J,P}\,V^{(J,P)}(\sqrt{s}\,)\, {\mathcal Y}^{(J,P)}_{\mu
\nu}(\bar q, q,w) \,,
\nonumber\\
&& {\mathcal Y}^{(3/2, -)}_{\mu \nu}(\bar q,q;w) =
\frac{1}{2}\left( g_{\mu \nu}-\frac{w_\mu \,w_\nu}{w^2}\right)
\left(1 + \frac{\wslash }{\sqrt{w^2}}\right)
\nonumber\\
&& \qquad \qquad \qquad \quad -\frac{1}{6} \left(\gamma_\mu
-\frac{w_\mu}{w^2}\,\wslash \right) \left(1 - \frac{\wslash
}{\sqrt{w^2}}\right) \left(\gamma_\nu -\frac{w_\nu}{w^2}\,\wslash
\right) \;. \label{def-proj}
\end{eqnarray}
where we recall the projector relevant for s-wave  scattering. At
leading order the effective scattering kernel $V_{\mu \nu}(\bar
k,k;w)$ may be identified with the on-shell projected
Bethe-Salpeter kernel $K_{\mu \nu} (\bar k,k;w)$. The merit of the
projectors is that they decouple the Bethe-Salpeter equation
(\ref{BS-coupled}) into orthogonal sectors labelled by the total
angular momentum $J$. Here we suppress an additional matrix
structure that follows since for given parity and total angular
momentum, $J \geq 3/2$, two distinct angular momentum states
couple. In general, for given $J$, the projector form a $2\times
2$ matrix, for which we displayed in (\ref{def-proj}) only its
leading $11$-component. The effect of the remaining components is
phase-space suppressed and not considered here.

In this work we neglect the t-dependence of the interaction kernel
insisting on $t=0$ in (\ref{def-K}). Following
\cite{LWF02,Copenhagen} the s-wave projected effective scattering
kernel, $V^{(I,S,C)}(\sqrt{s}\,)$, is readily constructed:
\begin{eqnarray}
V^{(I,S,C)}(\sqrt{s}\,) = \!\!\sum_{V\in [16]}\frac{C^{(I,S,C)}_V}{8\,m_V^2}\, \Big(
2\,\sqrt{s}-M-\bar M +(\bar M-M)\,\frac{\bar m^2 -m^2}{m_V^2} \Big) \,,
\label{VWT}
\end{eqnarray}
where $M$, $\bar M$ and $m, \bar m$ are the masses of initial and
final baryon and meson states. In (\ref{VWT}) and below we
suppress the reference to the angular momentum and parity $J^P=
\frac{3}{2}^-$. The partial-wave scattering amplitudes,
$M^{(I,S,C)}(\sqrt{s}\,)$, take the simple form
\begin{eqnarray}
&&  M^{(I,S,C)}(\sqrt{s}\,) = \Big[ 1- V^{(I,S,C)}(\sqrt{s}\,)\,J^{(I,S,C)}(\sqrt{s}\,)\Big]^{-1}\,
V^{(I,S,C)}(\sqrt{s}\,)\,.
\label{final-t}
\end{eqnarray}
The unitarity loop function, $J^{(I,S,C)}(\sqrt{s}\,)$, is a diagonal matrix. Each element
depends on the masses of intermediate meson and baryon, $m$ and $M$, respectively:
\begin{eqnarray}
&& J(\sqrt{s}\,) =
\Big(E+M\Big) \left( \frac59 + \frac{2\,E}{9\,M} + \frac{2\,E^2}{9\,M^2} \right)\,
\Big(I(\sqrt{s}\,)-I(\mu) \Big)\,,
\nonumber\\
\nonumber\\
&& I(\sqrt{s}\,)=\frac{1}{16\,\pi^2}
\left( \frac{p_{\rm cm}}{\sqrt{s}}\,
\left( \ln \left(1-\frac{s-2\,p_{\rm cm}\,\sqrt{s}}{m^2+M^2} \right)
-\ln \left(1-\frac{s+2\,p_{\rm cm}\sqrt{s}}{m^2+M^2} \right)\right)
\right.
\nonumber\\
&&\qquad \qquad + \left.
\left(\frac{1}{2}\,\frac{m^2+M^2}{m^2-M^2}
-\frac{m^2-M^2}{2\,s}
\right)
\,\ln \left( \frac{m^2}{M^2}\right) +1 \right)+I(0)\;,
\label{i-def}
\end{eqnarray}
where $\sqrt{s}= \sqrt{M^2+p_{\rm cm}^2}+ \sqrt{m^2+p_{\rm cm}^2}$
and $E= \sqrt{M^2+p_{\rm cm}^2}$ \cite{LWF02}. The finite widths of the
$\Delta (1232)$ and $\Sigma(1385)$ states is taken into account by folding the loop functions $J(\sqrt{s})$ of
(\ref{i-def}) with normalized spectral functions (see \cite{LWF02}).

A crucial ingredient of the approach developed in
\cite{LK02,LK04-axial} is its approximate crossing symmetry
guaranteed by a matching scheme. The matching scale $\mu $ in
(\ref{i-def}) depends on the quantum number $(I,S,C)$ but should
be chosen uniformly within a given sector. It also must be
independent on the $J$ and $P$
\cite{LK02,LK04-axial,Granada,Copenhagen,Hofmann:Lutz:2005}. We
insist on
\begin{eqnarray}
&& \mu = \sqrt{  m_{\rm th}^2+M_{\rm th}^2}\,,  \label{mu-def}
\end{eqnarray}
where $m_{\rm th}+M_{\rm th}$ is the mass of the lightest hadronic
channel. Note that almost all matching scales were encountered
already in \cite{Hofmann:Lutz:2005} when studying s-wave
resonances. As a consequence of (\ref{mu-def}) the s-channel and
u-channel unitarized amplitudes involving the lightest channels
can be matched smoothly at the matching point $\mu$
\cite{LK02,LK04-axial,Granada,Copenhagen}. The construction
(\ref{mu-def}) implies in addition that the effect of heavy
channels on the light channels is  suppressed naturally
\cite{Hofmann:Lutz:2005}.

\clearpage

\section{Numerical results}

In order to explore the formation of baryon resonances we study
generalized speed functions
\cite{Hoehler:speed,LK04-axial,Hofmann:Lutz:2005} of the simple
form
\begin{eqnarray}
{\rm Speed}_{ab}(\sqrt{s}\,) =
\Big| \frac{d}{ d \sqrt{s}\,}\,\big[\,M_{ab}(\sqrt{s}\,) \big] \Big|\,.
 \label{def-gen-speed}
\end{eqnarray}
If a partial-wave scattering amplitude develops a
resonance or bound state, close to that structure it may be approximated by
a pole and a background term. We write
\begin{eqnarray}
&& M_{ab}(\sqrt{s}\,) \simeq  -\frac{g^*_a\,g^{}_b}{\sqrt{s}-M_R+i\,\Gamma_R/2} + b_{ab} \,,
\label{Breit-Wigner}
\end{eqnarray}
with the resonance mass $M_R$ and width $\Gamma_R$. The dimension less
coupling constants $g_b$ and $g_a$  parameterize the coupling strength of
the resonance to the initial and final channels. The background term $b_{ab}$
is in general a complex number. If the scattering amplitude  has the form (\ref{Breit-Wigner})
its speed takes a maximum at the resonance mass $M_R$. The ratio of coupling constants to
total decay width $\Gamma_R$ is then determined by the value the speed takes at its maximum
\begin{eqnarray}
{\rm Speed}_{aa}(M_R) = \Bigg|\frac{2\,g_a}{\Gamma_R} \Bigg|^2 \,.
\label{def-speed}
\end{eqnarray}
We determine the resonance position and coupling constants by adjusting the parameters
$M_R$ and $g_a$ to the Speed of its associated amplitudes. This is an approximate
procedure, fully sufficient in view of the schematic nature of the computation.

It should be stressed that the properties of the resonance states
presented in the following ( see Tabs.
\ref{tab:charm-1}-\ref{tab:charm3} ) are subject to changes
expected when incorporating additional channels involving the
$1^-$ mesons and $\frac{1}{2}^+$ baryons. The latter channels are
required to arrive at results that are consistent with the
heavy-quark symmetry.

\subsection{$J^P=\frac{3}{2}^-$ resonances with charm minus one}

We begin with a presentation of results obtained for d-wave
resonances with negative charm. The possible existence of such
states was discussed first by Gignoux, Silvestre-Brac and Richard
twenty years ago
\cite{Gignoux:Silvestre-Brac:Richard:87,Lipkin:87}. Such states
were studied in more detail within a quark model
\cite{Genovese:98}, however, predicting that they are unbound.
Recently the H1 collaboration reported on a possible resonance
structure in the $D_-^* p$ invariant mass at 3.099 GeV, which
suggests an exotic state with negative charm, isospin one and zero
strangeness \cite{H1Collaboration}. This triggered various
theoretical investigations
\cite{Stewart:04,Kim:04,Sarac:05,Wessling:04,Pirjol:Schat:05,Bicudo:05}.
According to a QCD sum rule studies \cite{Kim:04,Sarac:05} the H1
signal is unlikely to be an s- or  p-wave resonance. The $J^P=
\frac{1}{2}^+$ state suggested in \cite{Kim:04,Sarac:05} has a
mass below 3 GeV. It should be mentioned that the H1 signal was
not confirmed by the FOCUS collaboration \cite{FOCUS
Collaboration}.

\begin{table}[t]
\rescale \setlength{\tabcolsep}{1.2mm}
\setlength{\arraycolsep}{2.2mm}
\renewcommand{\arraystretch}{0.75}
\begin{center}
\begin{tabular}{|ll|c|c|c|}
\hline $C=-1:$ &$ (\,I,\phantom{+}S)$  & $\rm state$ &
$\begin{array}{c} M_R [\rm MeV]  \\ \Gamma_R \,[\rm MeV] \end{array}$ & $|g_R|$  \\
\hline \hline &$(1,+0)$  & $\begin{array}{l}  \rule{0pt}{0.3ex}
\bar D \,\Delta
\end{array}$
& $\begin{array}{c} 2867 \\ 0 \end{array}$ & $\begin{array}{c} 5.8 \end{array}$ \\
\hline &$(\frac12,-1)$    & $
\begin{array}{l}
\bar{D} \,\Sigma
\end{array}$
& $\begin{array}{c} 2971 \\ 0  \end{array}$ & $\begin{array}{c} 5.9 \end{array}$ \\
\hline &$(\frac32,-1)$    & $
\begin{array}{l}
\bar D_s \Delta \\ \bar D \, \Sigma
\end{array}$
& $\begin{array}{c} 3098 \\ 0  \end{array}$ & $\begin{array}{c} 3.9 \\ 2.0 \end{array}$ \\
\hline &$(0,-2)$    & $
\begin{array}{l}
\bar D\, \Xi
\end{array}$
& $\begin{array}{c} 3045 \\ 0  \end{array}$ & $\begin{array}{c} 5.8 \end{array}$ \\
\hline &$(1,-2)$    & $
\begin{array}{l}
\bar D_s\Sigma \\ \bar D\, \Xi
\end{array}$
& $\begin{array}{c} 3170 \\ 0  \end{array}$ & $\begin{array}{c} 3.9 \\ 3.2 \end{array}$ \\
\hline &$(\frac12,-3)$    & $
\begin{array}{l}
\bar{D}_s \,\Xi \\ \bar D \, \Omega
\end{array}$
& $\begin{array}{c} 3211 \\ 0  \end{array}$ & $\begin{array}{c} 2.9 \\ 4.5 \end{array}$ \\
\hline
\end{tabular}
\caption{Spectrum of $J^P=\frac{3}{2}^-$ baryons with charm minus
one. } \label{tab:charm-1}
\end{center}
\end{table}

The properties of the $C=-1$ states as generated by the
coupled-channel equations (\ref{VWT}, \ref{final-t}, \ref{i-def})
are collected in Tab. \ref{tab:charm-1}. The spectrum is computed
insisting on the chiral relations (\ref{chiral-constraint-mesons},
\ref{chiral-constraint-baryons}) together with the leading order
large-$N_c$ relations (\ref{OZI-constraint},
\ref{OZI-constraint-baryons}). Relying on the KSFR relation
(\ref{KSFR-relation}) the binding energies are determined by the
universal vector coupling constant for which we take the value $g
=6.6$ from \cite{Hofmann:Lutz:2005}. It is emphasized that none of
the coupling constants (\ref{meson-SU4-result},
\ref{baryon-SU4-result})  that are estimated by an SU(4) ansatz
are relevant for the spectrum. Various molecules are formed. Their
SU(3) multiplet structure is readily worked out:
\begin{eqnarray}
&& 3 \otimes 10 = 15_1 \oplus 15_2 \,,
\nonumber\\
&& [15_1] \ni \left(
\begin{array}{c}
(1,+0)\\
(\frac{1}{2},-1),(\frac{3}{2},-1)\\
(0,-2),(1,-2)\\
(\frac{1}{2},-3)
\end{array}\right)\,,\qquad
[15_2] \ni \left(
\begin{array}{c}
(2,+0)\\
(\frac{3}{2},-1)\\
(1,-2)\\
(\frac{1}{2},-3) \\
(0,-4)
\end{array}\right)\,.
\label{}
\end{eqnarray}
Attraction is predicted in the first 15-plet and repulsion in the
second 15-plet. Summing the coefficient matrix $C_V$ over the nine
light vector mesons we obtain:
\begin{eqnarray}
\frac{1}{4\,g^2}\,\sum_{V \in [9]} C_V^{(C=-1)} =
C_{[15_1]}-3\,C_{[15_2]}\,. \label{charm+1-multiplet}
\end{eqnarray}
The strengths in the two multiplets reflect the coefficient in
front of the 140- and 120-plet of the SU(4) decomposition in
(\ref{SU4-decomposition}). This follows since charm-exchange
does not contribute here. The algebra
(\ref{charm+1-multiplet}) is directly reflected in Tab.
\ref{tab:charm-1}, which collects the masses and coupling
constants of the bound 15-plet. The SU(3) breaking pattern in the
multiplet is significant. The lightest state with $(I,S)=(1,0)$
comes at 2867 MeV, the heaviest state with $(I,S)=(1/2,-3)$ a mass
of 3211 MeV. So far the states predicted in Tab.
\ref{tab:charm-1} have not been observed. Since the $(1,0)$ state
has a mass below the $D\,N$ mass we do not see any indication that
the H1 signal could possibly be a result of a d-wave state.

\subsection{$J^P=\frac{3}{2}^-$ resonances with zero charm}

We turn to the resonances with $C=0$. The spectrum  falls into two
types of states. Resonances with masses above 3 GeV couple
strongly to mesons with non-zero charm content. In the SU(3) limit
those states form an octet. All other states have masses below 2.5
GeV. In the SU(3) limit they group into an octet and decuplet. The
presence of the heavy channels does not affect that part of the
spectrum at all. We reproduce the previous coupled-channel
computation \cite{Copenhagen}. The effect of using a spectral
distribution for the $\Delta(1232)$ and $\Sigma(1385)$ does not
alter the spectrum qualitatively \cite{Oset-remake}. There is
nothing we want to add to those results at this stage.

Most spectacular are the resonances with hidden charm above 3 GeV.
The multiplet structure of such states is readily understood. The
mesons with $C=-1$  form a triplet which is scattered off the
charmed baryons forming a sextet. We decompose the product into
irreducible tensors
\begin{eqnarray}
3 \otimes 6 = 8\oplus 10\,. \label{3times6}
\end{eqnarray}
The interaction is attractive in the crypto-exotic octet and
repulsive in the decuplet. For the formation of the states the charm-exchange
processes are irrelevant. This holds as long as the SU(4) estimates
(\ref{meson-SU4-result}, \ref{baryon-SU4-result}) give the coupling constants within
a factor of three. Thus the crypto-exotic sector may be characterized by the decomposition
\begin{eqnarray}
\frac{1}{4\,g^2}\,\sum_{V\in [9]}\,C^{(C=0)}_V =
C^{\rm crypto}_{[8]} - 2\, C^{\rm crypto}_{[10]}\,,
\label{crypto-decomposition}
\end{eqnarray}
where we assume the chiral and large-$N_c$ relations
(\ref{chiral-constraint-mesons}, \ref{chiral-constraint-baryons}) and (\ref{OZI-constraint},
\ref{OZI-constraint-baryons}).
The binding energies of the crypto-exotic states are large.
This is in part due to the large masses of the coupled-channel states: the kinetic
energy the attractive t-channel force has to overcome is reduced.
A second kinematical effect, which  further increases the binding
energy, is implied by the specific form of the t-channel exchange
(\ref{VWT}). It provides the factor $2\,\sqrt{s}-M-\bar M$. If
evaluated at threshold it scales with the meson mass.
We emphasize that the results are quite stable with respect to small variations of
the matching scale $\mu$ introduced
in (\ref{mu-def}): if we lower or increase $\mu$ by 20$\%$ the binding energies of the
crypto exotic states change by less than 20 MeV.

\begin{table}[t]
\rescale \setlength{\tabcolsep}{1.2mm}
\setlength{\arraycolsep}{2.2mm}
\renewcommand{\arraystretch}{0.75}
\begin{center}
\begin{tabular}{|ll|c|c|c|}
\hline $C=0:$ &$ (\,I,\phantom{+}S)$  & $\rm state$ &
$\begin{array}{c} M_R [\rm MeV]  \\ \Gamma_R \,[\rm MeV] \end{array}$ & $|g_R|$  \\
\hline \hline &$(\frac12,\phantom{+}0)$   &
$\begin{array}{l} \pi\, \Delta \\ K\,\Sigma \\ \bar D\,\Sigma_c \end{array}$
& $\begin{array}{c} 3430 \\ 0.50 \end{array}$ & $\begin{array}{c} 0.05  \\ 0.04  \\ 5.6 \end{array}$ \\
\hline
&$(0,-1)$    &
$\begin{array}{l} \pi\, \Sigma \\ K \,\Xi \\ \bar D\,\Xi_c \end{array}$
& $\begin{array}{c} 3538 \\ 0.63 \end{array}$ & $\begin{array}{c} 0.04 \\ 0.05 \\ 5.5 \end{array}$ \\
\hline
&$(1,-1)$    &
$\begin{array}{l} \pi\,\Sigma \\ \bar K \,\Delta \\ \eta \,\Sigma \\ K\, \Xi \\\eta' \Sigma \\ \eta_c \Sigma \\ \bar{D}_s \Sigma_c \\ \bar{D}\, \Xi_c \end{array}$
& $\begin{array}{c} 3720 \\ 0.83 \end{array}$
& $\begin{array}{c} 0.02 \\ 0.04 \\ 0.04 \\ 0.03 \\ 0.01 \\ 0.20 \\ 4.5 \\ 2.8 \end{array}$ \\
\hline
&$(\frac12,-2)$    &
$\begin{array}{l} \pi\, \Xi \\ \bar K\, \Sigma \\ \eta\, \Xi \\ K\, \Omega \\ \eta' \Xi \\ \eta_c \Xi \\ \bar D_s \Xi_c \\ \bar D\, \Omega_c \end{array}$
& $\begin{array}{c} 3742 \\ 1.1 \end{array}$
& $\begin{array}{c} 0.02 \\ 0.03 \\ 0.04 \\ 0.06 \\ 0.03 \\ 0.16 \\ 3.2 \\ 4.2  \end{array}$ \\
\hline
\end{tabular}
\caption{ Spectrum of crypto-exotic $J^P=\frac{3}{2}^-$ baryons with charm
zero.} \label{tab:charm0}
\end{center}
\end{table}

In Tab. \ref{tab:charm0} the spectrum of crypto-exotic baryons is
shown. The charm-exchange contributions are estimated by the SU(4)
ansatz (\ref{meson-SU4-result},  \ref{baryon-SU4-result}). For the
universal vector coupling constant we use  $g =6.6$  as before
\cite{Hofmann:Lutz:2005}. The octet of states is narrow as a
result of the OZI rule. The mechanism is analogous to the one
explaining the long life time of the $J/\Psi$-meson. The precise
values of the width parameters are sensitive to the SU(4)
estimates. In contrast to our previous study of crypto-exotic
s-wave resonances \cite{Hofmann:Lutz:2005}, channels that involve
the $\eta'$ meson do not to play a special role in the d-wave
spectrum. This is in part a consequence that the $\eta'$ channels
decouple from the crypto-exotic sector in the SU(3) limit. It is
interesting to observe that a narrow nucleon resonance  is
predicted at 3.42 GeV. One may speculate that the crypto-exotic
resonance claimed at 3.52 GeV in \cite{Karnaukhov:91} may be a
d-wave state. Given the uncertainties of the claim
\cite{Karnaukhov:91} we refrain from fine tuning the model
parameters as to push up the $(\frac{1}{2},0)$ state.

The results collected in Tab. \ref{tab:charm0} are subject to
large uncertainties. The evaluation of the total
width as well as a more reliable estimate of the binding energies
of the crypto-exotic states requires the consideration of further
partial-wave contributions.  The large binding energy
obtained suggest a more detailed study that is based on
a more realistic interaction taking into account in particular the
finite masses of the t-channel exchange processes.

\clearpage

\subsection{$J^P=\frac{3}{2}^-$ resonances with charm one}

At present we know very little about open-charm d-wave resonances.
Only two states $\Lambda_c(2625)$ and $\Xi_c(2815)$ are quoted by
the Particle Data Group \cite{PDG04}. So far no quantum numbers
are determined experimentally. The assignments are based on a
quark-model bias. An additional state $\Sigma_c(2800)$ was
discovered recently by the BELLE collaboration \cite{BELLE-Sigma}.
We consider it as a candidate for a d-wave state.

In a previous coupled-channel computation the effect of
the Goldstone bosons as they scatter off the sextet of charmed baryons
with $J^P=\frac{3}{2}^+$ was studied
\cite{LK05}. We confirm the striking prediction which suggest the existence of
bound $\bar 3, 6,\overline{15}$ systems
where attraction is foreseen in the anti-triplet, sextet and $\overline{15}$-plet with
decreasing strength. This result is reproduced quantitatively if the charm-exchange reactions
are neglected. In this case the decomposition
\begin{eqnarray}
\frac{1}{4\,g^2}\,\sum_{V\in [9]}\,C^{(C=1)}_V =
5\, C^{\rm chiral}_{[\bar 3]} + 3\,C^{\rm chiral}_{[6]}
+C^{\rm chiral}_{[\overline{15}]} -2\, C^{\rm chiral}_{[24]}\,,
\label{chiral-decomposition}
\end{eqnarray}
holds, where the suffix 'chiral' indicates that the multiplets are realized with states that
involve Goldstone bosons. The chiral and large-$N_c$ relations
(\ref{chiral-constraint-mesons}, \ref{chiral-constraint-baryons}) and (\ref{OZI-constraint},
\ref{OZI-constraint-baryons}) are assumed in (\ref{chiral-decomposition}).

\begin{table}[t]
\rescale \setlength{\tabcolsep}{1.2mm}
\setlength{\arraycolsep}{2.2mm}
\renewcommand{\arraystretch}{0.75}
\begin{center}
\begin{tabular}{|ll|c|c|c|c|c|}
\hline $C=1:$ &$ (\,I,\phantom{+}S)$  & $\rm state$ &
$\begin{array}{c} M_R [\rm MeV]  \\ \Gamma_R \,[\rm MeV]
\end{array}$ &
$|g_R|$  & $\begin{array}{c} M_R [\rm MeV]  \\ \Gamma_R \,[\rm MeV]  \end{array}$ & $|g_R|$  \\
\hline \hline
&$(\frac12,\phantom{+}1)$    & $\begin{array}{l} K\, \Sigma_c \end{array}$
& $\begin{array}{c} 2988 \\ 2.0 \end{array}$
& $\begin{array}{c} 1.6  \end{array}$
& $\begin{array}{c} -- \end{array}$ & \\
\hline
&$(0,\phantom{+}0)$
& $\begin{array}{l} \pi \Sigma_c \\ K\, \Xi_c \\ \bar D\,\Xi_{cc} \end{array}$
& $\begin{array}{c} 2660 \\ 53 \end{array}$
& $\begin{array}{c} 2.3 \\ 0.5 \\ 0.1  \end{array}$
& $\begin{array}{c} 3100 \\ 17 \end{array}$
& $\begin{array}{c} 0.50 \\ 2.3 \\ 0.08  \end{array}$\\
\cline{3-7}
& $(0,\phantom{+}0)$
& $\begin{array}{l} \pi \cdot \Sigma_c \\ K\, \Xi_c \\ \bar D\,\Xi_{cc} \end{array}$
& $\begin{array}{c} 4268 \\ 1.4 \end{array}$
& $\begin{array}{c} 0.08 \\ 0.06 \\ 4.9  \end{array}$
& $\begin{array}{c} -- \end{array}$ & \\
\hline
&$(1,\phantom{+}0)$
& $\begin{array}{l} \pi\,\Sigma_c \\ \eta\, \Sigma_c \\ D\, \Delta \\ K \, \Xi_c \\ D_s\Sigma \\ \eta' \Sigma_c \\ \bar D \, \Xi_{cc} \\ \eta_c \Sigma_c \end{array}$
& $\begin{array}{c} 2613 \\ 0 \end{array}$
& $\begin{array}{c} 0.05 \\ 0.03 \\ 8.6 \\ 0.01 \\ 3.1 \\ 0.09 \\ 0.01 \\ 0.74  \end{array}$
& $\begin{array}{c} 2716 \\ 128 \end{array}$
& $\begin{array}{c} 2.2  \\ 0.1  \\ 0.1 \\ 0.9  \\ 0.1 \\ 0.0  \\ 0.3  \\ 0.05  \end{array}$\\
\cline{3-7}
&$(1,\phantom{+}0)$
& $\begin{array}{l} \pi\,\Sigma_c \\ \eta\, \Sigma_c \\ D\, \Delta \\ K \, \Xi_c \\ D_s\Sigma \\ \eta' \Sigma_c \\ \bar D \, \Xi_{cc} \\ \eta_c \Sigma_c \end{array}$
& $\begin{array}{c} 3064 \\ 0 \end{array}$
& $\begin{array}{c} 0.44 \\ 1.4 \\ 0 \\ 1.2 \\ 0.08 \\ 0.0 \\ 0.18 \\ 0.03  \end{array}$
& $\begin{array}{c} -- \end{array}$ & \\
\hline
&$(\frac12,-1)$
& $\begin{array}{l} \pi \,\Xi_c \\ \bar K \, \Sigma_c \\ \eta\,\Xi_c \\ D\, \Sigma \\ K\,\Omega_c \\ D_s \Xi \\ \eta' \Xi_c \\ \bar D\, \Omega_{cc} \\ \bar D_s \Xi_{cc} \\ \eta_c\Xi_c \end{array}$
& $\begin{array}{c} 2762 \\ 0 \end{array}$
& $\begin{array}{c} 0.03 \\ 0.13 \\ 0.01 \\ 7.4 \\ 0.01 \\ 4.4 \\ 0.09 \\ 0.01 \\ 0.0 \\ 0.73  \end{array}$
& $\begin{array}{c} 2838 \\ 16 \end{array}$
& $\begin{array}{c} 1.0 \\ 3.3 \\ 1.3 \\ 0.2 \\ 0.43 \\ 0.2 \\ 0.0 \\ 0.05 \\ 0.30 \\ 0.05  \end{array}$\\
\cline{3-7}
&$(\frac12,-1)$
& $\begin{array}{l} \pi \,\Xi_c \\ \bar K \, \Sigma_c \\ \eta\,\Xi_c \\ D\, \Sigma \\ K\,\Omega_c \\ D_s \Xi \\ \eta' \Xi_c \\ \bar D\, \Omega_{cc} \\ \bar D_s \Xi_{cc} \\ \eta_c\Xi_c \end{array}$
& $\begin{array}{c} 3180 \\ 43 \end{array}$
& $\begin{array}{c} 0.76 \\ 0.37 \\ 1.2 \\ 0 \\ 2.1 \\ 0.1 \\ 0 \\ 0.15 \\ 0.06 \\ 0.03 \end{array}$
& $\begin{array}{c} 4468 \\ 3.3 \end{array}$
& $\begin{array}{c} 0.04 \\ 0.06 \\ 0.06 \\ 0.01 \\ 0.08 \\ 0.01 \\ 0.09 \\ 3.4 \\ 3.4 \\ 0.12  \end{array}$\\
\hline
&$(0,-2)$
& $\begin{array}{l} \bar K\, \Xi_c \\ \eta\, \Omega_c \\ D\, \Xi \\ D_s \Omega \\ \eta' \Omega_c \\ \bar D_s \Omega_{cc} \\ \eta_c \Omega_c \end{array}$
& $\begin{array}{c} 2843 \\ 0 \end{array}$
& $\begin{array}{c} 0.12 \\ 0.03 \\ 6.2 \\ 5.5 \\ 0.09 \\ 0.01  \\ 0.76  \end{array}$
& $\begin{array}{c} 3008 \\ 0 \end{array}$
& $\begin{array}{c} 2.5 \\ 1.8 \\ 0.14 \\ 0.25 \\ 0.01 \\ 0.37 \\ 0.06  \end{array}$\\
\hline \hline
\end{tabular}
\caption{Spectrum of $J^P=\frac{3}{2}^-$ resonances with charm one.} \label{tab:charm1}
\end{center}
\end{table}

Further multiplets are generated by the scattering of the anti-triplet mesons of the decuplet
baryons. We derive
\begin{eqnarray}
\frac{1}{4\,g^2}\,\sum_{V\in [9]}\,C^{(C=1)}_V =
5\, C^{\rm heavy}_{[6]} \,,
\label{heavy-decomposition}
\end{eqnarray}
where we use the suffix 'heavy' to indicate that the multiplets are formed by states involving the
D mesons. In Tab. \ref{tab:charm1} the spectrum of resonances is displayed where the charm-exchange
contributions are considered based on the SU(4) estimates
(\ref{meson-SU4-result}, \ref{baryon-SU4-result}). It should again be emphasized
that the amount of binding predicted is insensitive to the SU(4) estimates. The latter is a consequence of
the universally coupled light vector mesons.

The multiplet structure anticipated in (\ref{chiral-decomposition}, \ref{heavy-decomposition})
is clearly reflected in the spectrum.
For the readers'
convenience we recall the $(I,S)$ content of the various SU(3) multiplets:
\begin{eqnarray}
&& [\bar 3] \ni \left( \begin{array}{c} (0,0) \\ (\frac{1}{2},-1) \end{array}\right)\,, \qquad
[6] \ni \left(
\begin{array}{c}  (1,0) \\ (\frac{1}{2},-1) \\
(0,-2)\end{array}
\right) \,,\qquad
\nonumber\\
&& [\overline{15}] \ni
\left(
\begin{array}{c}
(\frac{1}{2},+1)\\
(0,0),(1,0)\\
(\frac{1}{2},-1),(\frac{3}{2},-1)\\
(1,-2)
\end{array}\right)\,,\qquad
 [24] \ni
\left(
 \begin{array}{c}
(\frac{3}{2},+1)\\
(1,0),(2,0)\\
(\frac{1}{2},-1),(\frac{3}{2},-1)\\
(0,-2),(1,-2)\\
(\frac{1}{2},-3)
\end{array}
\right) \,.
\label{recall-multiplets}
\end{eqnarray}
Strongest binding is predicted for the sextet states with
$\Sigma_c(2613)$, $\Xi_c(2762)$ and $\Omega_c(2843)$. Such states
are difficult to detect empirically since they couple only very
weakly to open channels involving Goldstone bosons. As a
consequence the widths of those states is typically below 1 MeV.

The spectrum of chiral excitations is much richer and presumably
easier to confirm experimentally. A triplet and sextet of states
is formed. The isospin singlet of the triplet at mass 2659 MeV
should be identified with the $\Lambda_c(2625)$. We underestimate
the binding somewhat. The isospin doublet of the triplet comes at
2838 MeV. Its small width of about 16 MeV only reflects its small
coupling strength to the $\pi \,\Xi_c$ channel. In \cite{LK05} the
latter state was identified with the $\Xi_c(2815)$ of the Particle
Data Group \cite{PDG04}. The attraction predicted in the chiral
sextet channels leads to additional resonances $\Sigma_c(2716)$,
$\Xi_c(3180)$ and $\Omega_c(3008)$. In \cite{LK05} the isospin
triplet state was suggested as a candidate for Belle's state at
2800 MeV \cite{BELLE-Sigma}. However, the identification is
troublesome due to a too large width.

Crypto-exotic states with $cc\bar c$ content are
formed by the scattering of the $3$-plet mesons with $C=-1$ off the triplet baryons with $C=2$:
\begin{eqnarray}
\frac{1}{4\,g^2}\,\sum_{V\in [9]}\,C^{(C=1)}_V =
C^{\rm crypto}_{[\bar 3]} - C^{\rm crypto}_{[6]}\,,
\label{crypto-decomposition-charm1}
\end{eqnarray}
where we predict strong attraction in the anti-triplet sector
only. The associated narrow states have masses 4277 MeV and 4491
MeV. The binding energies and widths of these states are expected
to be quite model dependent and more detailed
studies are required.

\subsection{$J^P=\frac{3}{2}^-$ resonances with charm two}

Double-charm baryon systems are very poorly understood at present.
There is a single published isospin doublet state claimed by the
SELEX collaboration at 3519 MeV \cite{SELEX} that carries zero
strangeness. There are hints that this state can not be the ground
state with $J^P=\frac{1}{2}^+ $ quantum numbers \cite{SELEX:Russ}.
We assigned it $J^P=\frac{3}{2}^+$ quantum numbers.

There are two types of molecules formed in the coupled-channel
computations. The chiral excitations of the $\frac{3}{2}^+$
baryons with $C=2$ form a strongly bound triplet  and a less bound
sextet of resonance. This part of the spectrum is analogous to the
one of the chiral excitations of open-charm mesons: attraction is
predicted in  the triplet and anti-sextet sectors
\cite{KL04,HL04}:
\begin{eqnarray}
8 \otimes 3 =3 \oplus \bar 6 \oplus 15 \,, \qquad
\qquad \bar 3 \otimes 6 = 3 \oplus 15 \,.
\label{decomp-cc}
\end{eqnarray}
Further molecules are formed by the systems composed of open-charm mesons and open-charm baryons.
The multiplet decomposition for the sextet baryons is given also in (\ref{decomp-cc}).
Strong attraction is predicted again in the triplet only. In the SU(3) limit there is no interaction
in the crypto-exotic $\bar D\,\Omega_{ccc}$ sector if the charm-exchange reactions are neglected.
All together we have
\begin{eqnarray}
&& \frac{1}{4\,g^2}\,\sum_{V\in [9]}\,C^{(C=2)}_V =
3\, C^{\rm chiral}_{[3]} + C^{\rm chiral}_{[\bar 6]}-C^{\rm chiral}_{[15]}
 +4\,C^{\rm heavy}_{[3]} \,.
\label{decomposition-charm2}
\end{eqnarray}

\begin{table}[t]
\rescale \setlength{\tabcolsep}{1.2mm}
\setlength{\arraycolsep}{2.2mm}
\renewcommand{\arraystretch}{0.75}
\begin{center}
\begin{tabular}{|ll|c|c|c|c|c|}
\hline $C=2:$ &$ (\,I,\phantom{+}S)$  & $\rm state$ &
$\begin{array}{c} M_R [\rm MeV]  \\ \Gamma_R \,[\rm MeV] \end{array}$ &
$|g_R|$  & $\begin{array}{c} M_R [\rm MeV]  \\ \Gamma_R \,[\rm MeV]  \end{array}$ & $|g_R|$  \\
\hline
\hline
&$(0,\phantom{+}1)$
& $\begin{array}{l} K \,\Xi_{cc} \end{array}$
& $\begin{array}{c} 3983 \\ 0 \end{array}$ & $\begin{array}{c} 2.1 \end{array}$
& $\begin{array}{c} -- \end{array}$ & \\
\hline
&$(\frac12,\phantom{+}0)$
& $\begin{array}{l} \pi\,\Xi_{cc} \\ \eta\, \Xi_{cc} \\ K\, \Omega_{cc} \\ D\, \Sigma_c \\ \eta' \Xi_{cc} \\ D_s \Xi_c \\ \bar D\, \Omega_{ccc} \\ \eta_c \Omega_{cc} \end{array}$
& $\begin{array}{c} 3671 \\ 0.01 \end{array}$ & $\begin{array}{c} 0.10 \\ 0.04 \\ 0.03 \\ 6.4  \\ 0.13 \\ 2.7 \\ 0.02 \\ 0.90  \end{array}$
& $\begin{array}{c} 3723 \\ 148  \end{array}$ & $\begin{array}{c} 2.2  \\ 0.2  \\ 0.85 \\ 0.2  \\ 0    \\ 0.1 \\ 0.3  \\ 0.1   \end{array}$\\
\cline{3-7}
&$(\frac12,\phantom{+}0)$
& $\begin{array}{l} \pi\,\Xi_{cc} \\ \eta\, \Xi_{cc} \\ K\, \Omega_{cc} \\ D\, \Sigma_c \\ \eta' \Xi_{cc} \\ D_s \Xi_c \\ \bar D\, \Omega_{ccc} \\ \eta_c \Omega_{cc} \end{array}$
& $\begin{array}{c} 4046 \\ 19 \end{array}$
& $\begin{array}{c} 0.54 \\ 1.4 \\ 1.6 \\ 0.05 \\ 0.0 \\ 0.07 \\ 0.15 \\ 0.04 \end{array}$
& $\begin{array}{c} -- \end{array}$ & \\
\hline
&$(0,-1)$
& $\begin{array}{l} \bar K\, \Xi_{cc}\\ \eta\, \Omega_{cc} \\ D\,\Xi_c \\ \eta'\Omega_{cc} \\ D_s \Omega_c \\ \bar D_s \Omega_{ccc} \\ \eta_c\Omega_{cc} \end{array}$
& $\begin{array}{c} 3761 \\ 0 \end{array}$ & $\begin{array}{c} 0.21 \\ 0.04 \\ 5.6  \\ 0.14 \\ 4.0  \\ 0.0  \\ 1.0  \end{array}$
& $\begin{array}{c} 3863 \\ 0 \end{array}$ & $\begin{array}{c} 2.7  \\ 1.3  \\ 0.30 \\ 0.02 \\ 0.30 \\ 0.37 \\ 0.11 \end{array}$\\
\hline
\end{tabular}
\caption{Spectrum of $J^P=\frac{3}{2}^-$ baryons with charm two.}
\label{tab:charm2}
\end{center}
\end{table}

For the readers' convenience we recall the isospin strangeness content of the various multiplets:
\begin{eqnarray}
 && [3] \ni \left( \begin{array}{c} (\frac{1}{2},0) \\  (0,-1) \end{array}\right)\,, \quad \!
[\bar 6] \ni \left(
\begin{array}{c}  (0,+1) \\ (\frac{1}{2},0) \\
(1,-1)\end{array}
\right) \,, \quad \!
[15] \ni \left(
\begin{array}{c}  (1,+1) \\ (\frac{1}{2},0)\,,(\frac{3}{2},0) \\
(0,-1)\,,(1,-1)\\
(\frac{1}{2},-2)\end{array}
\right)\,.
\end{eqnarray}

In Tab. \ref{tab:charm2} the masses and coupling constants of the
double-charm molecules are collected. Strongest binding is
foreseen for the $\Xi_{cc}(3671)$ and $\Omega_{cc}(3761)$ states
which couple dominantly to the D mesons. Those resonances are
quite narrow since their decay into final states with Goldstone
bosons is suppressed. The relevant coupling constants reflect the
SU(4) estimates (\ref{meson-SU4-result}, \ref{baryon-SU4-result}).
A further triplet of double-charm molecules with $\Xi_{cc}(3723)$
and $\Omega_{cc}(3863)$ couple strongly to the final states with
Goldstone bosons. The isospin-doublet state is broad due to its
large coupling constant to the open $\pi\, \Xi_{cc}$ channel. The
weakly bound sextet of chiral excitations is identified most
clearly in the $(0,1)$ and $(\frac{1}{2},0)$ sectors: states of
mass 3983 MeV and 4046 MeV are predicted. Their $(1,+1)$ partner
is difficult to discriminate from the $\bar K\, \Xi_{cc}$
threshold and therefor not included in Tab. \ref{tab:charm2}. Due
to their shallow binding correction terms in the interaction may
or may not lead to their disappearance.

We expect the results for those resonances which couple strongly
to the final states with Goldstone bosons to be more reliable than
those for resonances which couple dominantly to the D mesons. For
instance in the $(\frac{1}{2},0)$ sector, one might wonder about
the effect of the s-wave $D^*\, \Sigma_c(2455)$ state, which is just
about 80 MeV higher in energy than the $D \,\Sigma_c(2520)$ state.
Heavy quark symmetry predicts great similarities, not only between
the charmed $0^-$ and $1^-$ mesons, but also between the charmed
baryons with $\frac{1}{2}^+$ and $\frac{3}{2}^+$. Thus a more reliable
computation asks for the inclusion of additional channels as to
arrive at results compatible with the constraints set by heavy-quark
symmetry.

\begin{table}[b]
\rescale \setlength{\tabcolsep}{1.2mm}
\setlength{\arraycolsep}{2.2mm}
\renewcommand{\arraystretch}{0.75}
\begin{center}
\begin{tabular}{|ll|c|c|c|c|c|}
\hline $C=3:$ &$ (\,I,\phantom{+}S)$  & $\rm state$ &
$\begin{array}{c} M_R [\rm MeV]  \\ \Gamma_R \,[\rm MeV] \end{array}$ & $|g_R|$  \\
\hline \hline &$(0,\phantom{+}0)$    & $\begin{array}{l} \eta\,
\Omega_{ccc} \\ D\,\Xi_{cc} \\ \eta' \Omega_{ccc} \\  D_s
\Omega_{cc} \\ \eta_c\Omega_{ccc}
\end{array}$
& $\begin{array}{c} 4369 \\ 0 \end{array}$ & $\begin{array}{c} 0.02 \\ 5.5 \\ 0.03 \\ 2.8 \\ 1.3  \end{array}$ \\
\hline
\end{tabular}
\caption{Spectrum of $J^P=\frac{3}{2}^-$ baryons with charm
three.} \label{tab:charm3}
\end{center}
\end{table}

\subsection{$J^P=\frac{3}{2}^-$ resonances with charm three}

We close the result section with a discussion of triple-charm baryon states. They are formed
by scattering the triplet baryons with $C=2$ of the anti-triplet mesons with $C=1$.
It holds
\begin{eqnarray}
&& \frac{1}{4\,g^2}\,\sum_{V\in [9]}\,C^{(C=3)}_V =3\, C^{\rm
heavy}_{[1]} \,, \label{decomposition-charm3}
\end{eqnarray}
where the chiral and large-$N_c$ relations
(\ref{chiral-constraint-mesons}, \ref{chiral-constraint-baryons})
and (\ref{OZI-constraint}, \ref{OZI-constraint-baryons}) are
assumed. The decomposition (\ref{decomposition-charm3}) reflects
the fact that Goldstone bosons do not interact with the SU(3)
singlet $\Omega_{ccc}$ at leading order in the chiral expansion.
In Tab. \ref{tab:charm3} we predict the properties of the bound
singlet state at mass 4369 MeV. This analogous to the s-wave
spectrum observed in \cite{Hofmann:Lutz:2005}.

\section{Summary}

We have performed a coupled-channel study of d-wave baryon
resonances with charm $-1,0,1,2,3,4$. A rich spectrum is predicted
in terms of a t-channel force defined by the exchange of light
vector mesons. All relevant coupling constants are obtained from
chiral and large-$N_c$ properties of QCD.  The results of this
work should be taken cautiously since it remains to study the
effect of additional terms in the interaction kernel. Moreover, in
order to restore the heavy-quark symmetry additional channels
involving $1^-$ and $\frac{1}{2}^+$ states have to be
incorporated.

Two distinct types of states are dynamically generated. A spectrum
of chiral excitations is formed due the interaction of the
Goldstone bosons with the baryon ground states \cite{KL04,LK05}.
Additional states are due to the coupled-channel dynamics of the D
mesons. Those states decouple from the spectrum of chiral
excitations to a large extent. The mixing of the two spectra
requires charm-exchange interactions that are suppressed
naturally. We estimated the poorly known strength of the
charm-exchange by making a SU(4) ansatz for three-point vertices.

Coupled-channel dynamics driven by the D mesons lead to a strongly
bound 15-plet of $C=-1$ states and  a narrow crypto-exotic octet
of charm-zero states. In the $C=+1$ and $C=+2$ sector a sextet and
triplet of narrow resonances is formed due to the interaction of D
mesons with the baryon decuplet and sextet respectively. A singlet
triple-charm state is foreseen below 4.5 GeV. We do not have a
good candidate for the possible $C=-1$ state of the H1
collaboration \cite{H1Collaboration}. Though we predict attraction
in the isospin triplet and strangeness zero sector, our state
below 2.9 GeV is too strongly bound. Most spectacular is the
prediction of narrow crypto-exotic baryons with charm zero forming
below 4 GeV. Such states contain a $c \bar c$ pair. Their widths
parameters are small due to the OZI rule, like it is the case for
the $J/\Psi$ meson. The narrow nucleon resonance close to 3.5 GeV
may be a natural candidate for the exotic signal claimed in
\cite{Karnaukhov:91}.

{\bfseries{Acknowledgments}}

M.F.M.L. acknowledges discussions with J.S. Lange.

\section{Appendix A}

\begin{table}
\rescale
\setlength{\tabcolsep}{1.2mm}
\setlength{\arraycolsep}{3.0mm}
\renewcommand{\arraystretch}{1.0}
\begin{center}
\begin{tabular}{|l|l|l|c|}
\hline
\makebox[1cm][l]{$(I,\phantom{x}S)$} & \makebox[1cm][l]{$(a,b)$} &
\makebox[1.cm][l]{$V$} & \makebox[9.8cm][l]{$C_{V,ab}^{(I,S,-1)}\qquad \qquad \quad $ charm minus one} \\
\hline
\hline
$(1,+0)$    &$(1,1)$    &$\rho$     & $\frac53\,g_{00}^9\,h_{\bar3\bar3}^9$ \\
        &       &$\omega$   & $8\,g_{00}^1\,h_{\bar3\bar3}^1 + 4\,g_{00}^9\,h_{\bar3\bar3}^1 + 2\,g_{00}^1\,h_{\bar3\bar3}^9 + g_{00}^9\,h_{\bar3\bar3}^9$ \\
        &       &$\phi$     & $4\,g_{00}^1\,h_{\bar3\bar3}^1 + 2\,g_{00}^1\,h_{\bar3\bar3}^9$ \\
        &       &$J/\Psi$   & $2\,g_{00}^0\,h_{\bar3\bar3}^0$ \\
\hline
\hline
$(2,+0)$    &$(1,1)$    &$\rho$     & $-g_{00}^9\,h_{\bar3\bar3}^9$ \\
        &       &$\omega$   & $8\,g_{00}^1\,h_{\bar3\bar3}^1 + 4\,g_{00}^9\,h_{\bar3\bar3}^1 + 2\,g_{00}^1\,h_{\bar3\bar3}^9 + g_{00}^9\,h_{\bar3\bar3}^9$ \\
        &       &$\phi$     & $4\,g_{00}^1\,h_{\bar3\bar3}^1 + 2\,g_{00}^1\,h_{\bar3\bar3}^9$ \\
\hline
\hline
\end{tabular}
\caption{The coupled-channel structure of the t-channel exchange in (\ref{def-K}).}
\label{tab:appendix:1}
\end{center}
\end{table}

\begin{table}
\rescale
\setlength{\tabcolsep}{1.2mm}
\setlength{\arraycolsep}{3.0mm}
\renewcommand{\arraystretch}{1.0}
\begin{center}
\begin{tabular}{|l|l|l|c|}
\hline
\makebox[1cm][l]{$(I,\phantom{x}S)$} & \makebox[1cm][l]{$(a,b)$} &
\makebox[1.cm][l]{$V$} & \makebox[9.8cm][l]{$C_{V,ab}^{(I,S,0)}\qquad \qquad \quad $ charm zero} \\
\hline
\hline
$(1,+1)$    &$(1,1)$    &$\rho$     & $\frac{10}3\,g_{00}^9\,h_{99}^9$ \\
        &       &$\omega$   & $-4\,g_{00}^1\,h_{99}^9 - 2\,g_{00}^9\,h_{99}^9$ \\
        &       &$\phi$     & $4\,g_{00}^1\,h_{99}^9$ \\
\hline
\hline
$(2,+1)$    &$(1,1)$    &$\rho$     & $-2\,g_{00}^9\,h_{99}^9$ \\
        &       &$\omega$   & $-4\,g_{00}^1\,h_{99}^9 - 2\,g_{00}^9\,h_{99}^9$ \\
        &       &$\phi$     & $4\,g_{00}^1\,h_{99}^9$ \\
\hline
\hline
$(\frac12,+0)$  &$(1,1)$    &$\rho$     & $\frac{20}3\,g_{00}^9\,h_{99}^9$ \\
\hline
\hline
\end{tabular}
\caption{The coupled-channel structure of the t-channel exchange in (\ref{def-K}).}
\label{tab:appendix:2}
\end{center}
\end{table}

\begin{table}
\rescale
\setlength{\tabcolsep}{1.2mm}
\setlength{\arraycolsep}{3.0mm}
\renewcommand{\arraystretch}{1.0}
\begin{center}
\begin{tabular}{|l|l|l|c|}
\hline
\makebox[1cm][l]{$(I,\phantom{x}S)$} & \makebox[1cm][l]{$(a,b)$} &
\makebox[1.cm][l]{$V$} & \makebox[9.8cm][l]{$C_{V,ab}^{(I,S,0)}\qquad \qquad \quad $ charm zero} \\
\hline
\hline
$(1,-1)$    &$(1,1)$    &$\rho$     & $\frac83\,g_{00}^9\,h_{99}^9$ \\
\cline{2-3}
        &$(1,2)$    &$K$        & $\frac43\,g_{00}^9\,h_{99}^9$ \\
\cline{2-3}
        &$(2,2)$    &$\rho$     & $\frac{10}3\,g_{00}^9\,h_{99}^9$ \\
        &       &$\omega$   & $4\,g_{00}^1\,h_{99}^9 + 2\,g_{00}^9\,h_{99}^9$ \\
        &       &$\phi$     & $-4\,g_{00}^1\,h_{99}^9$ \\
\cline{2-3}
        &$(2,3)$    &$K$        & $-\sqrt{\frac{32}{3}}\,g_{00}^9\,h_{99}^9$ \\
\cline{2-3}
        &$(1,4)$    &$K$        & $\frac83\,g_{00}^9\,h_{99}^9$ \\
\hline
\end{tabular}
\caption{The coupled-channel structure of the t-channel exchange in (\ref{def-K}).}
\label{tab:appendix:3}
\end{center}
\end{table}

\begin{table}
\rescale
\setlength{\tabcolsep}{1.2mm}
\setlength{\arraycolsep}{3.0mm}
\renewcommand{\arraystretch}{1.0}
\begin{center}
\begin{tabular}{|l|l|l|c|}
\hline
\makebox[1cm][l]{$(I,\phantom{x}S)$} & \makebox[1cm][l]{$(a,b)$} &
\makebox[1.cm][l]{$V$} & \makebox[9.8cm][l]{$C_{V,ab}^{(I,S,0)}\qquad \qquad \quad $ charm zero} \\
\hline
        &$(2,7)$    &$D^*$      & $\sqrt{2}\,g_{06}^{\bar3}\,h_{\bar39}^{\bar3} + \sqrt{2}\,g_{06}^{\bar3}\,h_{9\bar3}^{\bar3}$ \\
\cline{2-3}
        &$(3,7)$    &$D^*_s$    & $-\sqrt{\frac83}\,\,g_{06}^{\bar3}\,h_{\bar39}^{\bar3} - \sqrt{\frac83}\,\,g_{06}^{\bar3}\,h_{9\bar3}^{\bar3}$ \\
\cline{2-3}
        &$(5,7)$    &$D^*_s$    & $4\,g_{06}^{\bar3}\,h_{1\bar3}^{\bar3} + 4\,g_{06}^{\bar3}\,h_{\bar31}^{\bar3} - \frac43\,g_{06}^{\bar3}\,h_{\bar39}^{\bar3} - \frac43\,g_{06}^{\bar3}\,h_{9\bar3}^{\bar3}$ \\
\cline{2-3}
        &$(6,7)$    &$D^*_s$    & $-\sqrt{\frac{8}{3}}\,g_{06}^{\bar3}\,h_{0\bar3}^{\bar3} - \sqrt{\frac{8}{3}}\,g_{06}^{\bar3}\,h_{\bar30}^{\bar3}$ \\
\cline{2-3}
        &$(7,7)$    &$\omega$   & $8\,g_{66}^1\,h_{\bar3\bar3}^1 + 2\,g_{66}^9\,h_{\bar3\bar3}^1 + 4\,g_{66}^1\,h_{\bar3\bar3}^9 + g_{66}^9\,h_{\bar3\bar3}^9$ \\
        &       &$\phi$     & $4\,g_{66}^1\,h_{\bar3\bar3}^1 + 2\,g_{66}^9\,h_{\bar3\bar3}^1$ \\
        &       &$J/\Psi$   & $2\,g_{66}^0\,h_{\bar3\bar3}^0$ \\
\end{tabular}
\caption{The coupled-channel structure of the t-channel exchange in (\ref{def-K}).}
\label{tab:appendix:4}
\end{center}
\end{table}

\begin{table}
\rescale
\setlength{\tabcolsep}{1.2mm}
\setlength{\arraycolsep}{3.0mm}
\renewcommand{\arraystretch}{1.0}
\begin{center}
\begin{tabular}{|l|l|l|c|}
\hline
\makebox[1cm][l]{$(I,\phantom{x}S)$} & \makebox[1cm][l]{$(a,b)$} &
\makebox[1.cm][l]{$V$} & \makebox[9.8cm][l]{$C_{V,ab}^{(I,S,1)}\qquad \qquad \quad $ charm one} \\
\hline
\hline
$(\frac12,+1)$  &$(1,1)$    &$\rho$     & $4\,g_{66}^9\,h_{99}^9$ \\
        &       &$\omega$   & $-4\,g_{66}^1\,h_{99}^9 - 2\,g_{66}^9\,h_{99}^9$ \\
        &       &$\phi$     & $4\,g_{66}^1\,h_{99}^9$ \\
\hline
\hline
$(\frac32,+1)$  &$(1,1)$    &$\rho$     & $-2\,g_{66}^9\,h_{99}^9$ \\
        &       &$\omega$   & $-4\,g_{66}^1\,h_{99}^9 - 2\,g_{66}^9\,h_{99}^9$ \\
        &       &$\phi$     & $4\,g_{66}^1\,h_{99}^9$ \\
\cline{2-3}
        &$(1,2)$    &$D^*$      & $-\sqrt{2}\,g_{06}^{\bar3}\,h_{\bar39}^{\bar3} - \sqrt{2}\,g_{06}^{\bar3}\,h_{9\bar3}^{\bar3}$ \\
\hline
\end{tabular}
\caption{The coupled-channel structure of the t-channel exchange in (\ref{def-K}).}
\label{tab:appendix:5}
\end{center}
\end{table}

\begin{table}
\rescale
\setlength{\tabcolsep}{1.2mm}
\setlength{\arraycolsep}{3.0mm}
\renewcommand{\arraystretch}{1.0}
\begin{center}
\begin{tabular}{|l|l|l|c|}
\hline
\makebox[1cm][l]{$(I,\phantom{x}S)$} & \makebox[1cm][l]{$(a,b)$} &
\makebox[1.cm][l]{$V$} & \makebox[9.8cm][l]{$C_{V,ab}^{(I,S,1)}\qquad \qquad \quad $ charm one} \\
\hline
        &$(7,7)$    &$\rho$     & $\frac12\,g_{33}^9\,h_{\bar3\bar3}^9$ \\
        &       &$\omega$   & $8\,g_{33}^1\,h_{\bar3\bar3}^1 + 2\,g_{33}^9\,h_{\bar3\bar3}^1 + 2\,g_{33}^1\,h_{\bar3\bar3}^9 + \frac12\,g_{33}^9\,h_{\bar3\bar3}^9$ \\
        &       &$\phi$     & $4\,g_{33}^1\,h_{\bar3\bar3}^1 + 2\,g_{33}^9\,h_{\bar3\bar3}^1 + 2\,g_{33}^1\,h_{\bar3\bar3}^9 + g_{33}^9\,h_{\bar3\bar3}^9$ \\
        &       &$J/\Psi$   & $2\,g_{33}^0\,h_{\bar3\bar3}^0$ \\
\cline{2-3}
        &$(3,8)$    &$D^*$      & $\sqrt{\frac{16}3}\,g_{06}^{\bar3}\,h_{0\bar3}^{\bar3} + \sqrt{\frac{16}3}\,g_{06}^{\bar3}\,h_{\bar30}^{\bar3}$ \\
\cline{2-3}
        &$(5,8)$    &$D^*_s$    & $\sqrt{\frac43}\,g_{06}^{\bar3}\,h_{0\bar3}^{\bar3} + \sqrt{\frac43}\,g_{06}^{\bar3}\,h_{\bar30}^{\bar3}$ \\
\cline{2-3}
        &$(7,8)$    &$D^*$      & $\sqrt{2}\,g_{63}^{\bar3}\,h_{0\bar3}^{\bar3} + \sqrt{2}\,g_{63}^{\bar3}\,h_{\bar30}^{\bar3}$ \\
\hline
\end{tabular}
\caption{The coupled-channel structure of the t-channel exchange in (\ref{def-K}).}
\label{tab:appendix:6}
\end{center}
\end{table}

\begin{table}
\rescale
\setlength{\tabcolsep}{1.2mm}
\setlength{\arraycolsep}{3.0mm}
\renewcommand{\arraystretch}{1.0}
\begin{center}
\begin{tabular}{|l|l|l|c|}
\hline
\makebox[1cm][l]{$(I,\phantom{x}S)$} & \makebox[1cm][l]{$(a,b)$} &
\makebox[1.cm][l]{$V$} & \makebox[9.8cm][l]{$C_{V,ab}^{(I,S,1)}\qquad \qquad \quad $ charm one} \\
\hline
        &$(3,9)$    &$D^*_s$    & $-\sqrt{\frac{1}{3}}\,g_{63}^{\bar3}\,h_{\bar39}^{\bar3} - \sqrt{\frac{1}{3}}\,g_{63}^{\bar3}\,h_{9\bar3}^{\bar3}$ \\
\cline{2-3}
        &$(7,9)$    &$D^*_s$    & $\sqrt{6}\,g_{63}^{\bar3}\,h_{1\bar3}^{\bar3} + \sqrt{6}\,g_{63}^{\bar3}\,h_{\bar31}^{\bar3} - \sqrt{\frac{2}{3}}\,g_{63}^{\bar3}\,h_{\bar39}^{\bar3} - \sqrt{\frac{2}{3}}\,g_{63}^{\bar3}\,h_{9\bar3}^{\bar3}$ \\
\cline{2-3}
        &$(8,9)$    &$K$        & $g_{33}^9\,h_{\bar3\bar3}^9$ \\
\cline{2-3}
        &$(9,9)$    &$\omega$   & $8\,g_{33}^1\,h_{\bar3\bar3}^1 + 2\,g_{33}^9\,h_{\bar3\bar3}^1 + 4\,g_{33}^1\,h_{\bar3\bar3}^9 + g_{33}^9\,h_{\bar3\bar3}^9$ \\
        &       &$\phi$     & $4\,g_{33}^1\,h_{\bar3\bar3}^1 + 2\,g_{33}^9\,h_{\bar3\bar3}^1$ \\
        &       &$J/\Psi$   & $2\,g_{33}^0\,h_{\bar3\bar3}^0$ \\
\cline{2-3}
        &$(4,10)$   &$D^*$      & $2\,g_{06}^{\bar3}\,h_{0\bar3}^{\bar3} + 2\,g_{06}^{\bar3}\,h_{\bar30}^{\bar3}$ \\
\hline
\hline
\end{tabular}
\caption{The coupled-channel structure of the t-channel exchange in (\ref{def-K}).}
\label{tab:appendix:7}
\end{center}
\end{table}

\begin{table}
\rescale
\setlength{\tabcolsep}{1.2mm}
\setlength{\arraycolsep}{3.0mm}
\renewcommand{\arraystretch}{1.0}
\begin{center}
\begin{tabular}{|l|l|l|c|}
\hline
\makebox[1cm][l]{$(I,\phantom{x}S)$} & \makebox[1cm][l]{$(a,b)$} &
\makebox[1.cm][l]{$V$} & \makebox[9.8cm][l]{$C_{V,ab}^{(I,S,1)}\qquad \qquad \quad $ charm one} \\
\hline
$(0.-2)$&$(3,7)$    &$D^*$      & $\sqrt{\frac{8}{3}}\,g_{06}^{\bar3}\,h_{0\bar3}^{\bar3} + \sqrt{\frac{8}{3}}\,g_{06}^{\bar3}\,h_{\bar30}^{\bar3}$ \\
\cline{2-3}
        &$(4,7)$    &$D^*_s$    & $2\,g_{06}^{\bar3}\,h_{0\bar3}^{\bar3} + 2\,g_{06}^{\bar3}\,h_{\bar30}^{\bar3}$ \\
\cline{2-3}
        &$(6,7)$    &$D^*_s$    & $-\sqrt{2}\,g_{63}^{\bar3}\,h_{0\bar3}^{\bar3} - \sqrt{2}\,g_{63}^{\bar3}\,h_{\bar30}^{\bar3}$ \\
\hline
\hline
$(1,-2)$    &$(1,2)$    &$K$        & $-\sqrt{8}\,g_{66}^9\,h_{99}^9$ \\
\cline{2-3}
        &$(2,2)$    &$\rho$     & $-g_{66}^9\,h_{99}^9$ \\
        &       &$\omega$   & $4\,g_{66}^1\,h_{99}^9 + g_{66}^9\,h_{99}^9$ \\
        &       &$\phi$     & $-4\,g_{66}^1\,h_{99}^9 - 2\,g_{66}^9\,h_{99}^9$ \\
\hline
\end{tabular}
\caption{The coupled-channel structure of the t-channel exchange in (\ref{def-K}).}
\label{tab:appendix:8}
\end{center}
\end{table}

\begin{table}
\rescale
\setlength{\tabcolsep}{1.2mm}
\setlength{\arraycolsep}{3.0mm}
\renewcommand{\arraystretch}{1.0}
\begin{center}
\begin{tabular}{|l|l|l|c|}
\hline
\makebox[1cm][l]{$(I,\phantom{x}S)$} & \makebox[1cm][l]{$(a,b)$} &
\makebox[1.cm][l]{$V$} & \makebox[9.8cm][l]{$C_{V,ab}^{(I,S,2)}\qquad \qquad \quad $ charm two} \\
\hline
        &$(5,6)$    &$D^*_s$    & $-\sqrt{6}\,g_{63}^{\bar3}\,h_{1\bar3}^{\bar3} - \sqrt{6}\,g_{63}^{\bar3}\,h_{\bar31}^{\bar3} + \sqrt{\frac{2}{3}}\,g_{63}^{\bar3}\,h_{\bar39}^{\bar3} + \sqrt{\frac{2}{3}}\,g_{63}^{\bar3}\,h_{9\bar3}^{\bar3}$ \\
\cline{2-3}
        &$(6,6)$    &$\omega$   & $-8\,g_{66}^1\,h_{\bar3\bar3}^1 - 2\,g_{66}^9\,h_{\bar3\bar3}^1 - 4\,g_{66}^1\,h_{\bar3\bar3}^9 - g_{66}^9\,h_{\bar3\bar3}^9$ \\
        &       &$\phi$     & $-4\,g_{66}^1\,h_{\bar3\bar3}^1 - 2\,g_{66}^9\,h_{\bar3\bar3}^1$ \\
        &       &$J/\Psi$   & $-2\,g_{66}^0\,h_{\bar3\bar3}^0$ \\
\cline{2-3}
        &$(1,7)$    &$D^*$      & $-\sqrt{\frac{3}{2}}\,g_{31}^{\bar3}\,h_{\bar39}^{\bar3} - \sqrt{\frac{3}{2}}\,g_{31}^{\bar3}\,h_{9\bar3}^{\bar3}$ \\
\cline{2-3}
        &$(2,7)$    &$D^*$      & $-\sqrt{\frac{1}{6}}\,g_{31}^{\bar3}\,h_{\bar39}^{\bar3} - \sqrt{\frac{1}{6}}\,g_{31}^{\bar3}\,h_{9\bar3}^{\bar3}$ \\
\cline{2-3}
        &$(3,7)$    &$D^*_s$    & $-g_{31}^{\bar3}\,h_{\bar39}^{\bar3} - g_{31}^{\bar3}\,h_{9\bar3}^{\bar3}$ \\
\hline
\hline
\end{tabular}
\caption{The coupled-channel structure of the t-channel exchange in (\ref{def-K}).}
\label{tab:appendix:9}
\end{center}
\end{table}

\begin{table}
\rescale
\setlength{\tabcolsep}{1.2mm}
\setlength{\arraycolsep}{3.0mm}
\renewcommand{\arraystretch}{1.0}
\begin{center}
\begin{tabular}{|l|l|l|c|}
\hline
\makebox[1cm][l]{$(I,\phantom{x}S)$} & \makebox[1cm][l]{$(a,b)$} &
\makebox[1.cm][l]{$V$} & \makebox[9.8cm][l]{$C_{V,ab}^{(I,S,2)}\qquad \qquad \quad $ charm two} \\
\hline
\hline
$(1,-1)$        &$(1,2)$        &$K$            & $2\,g_{33}^9\,h_{99}^9$ \\
\cline{2-3}
                &$(2,2)$        &$\rho$         & $g_{33}^9\,h_{99}^9$ \\
                &               &$\omega$       & $4\,g_{33}^1\,h_{99}^9 + g_{33}^9\,h_{99}^9$ \\
                &               &$\phi$         & $-4\,g_{33}^1\,h_{99}^9 - 2\,g_{33}^9\,h_{99}^9$ \\
\hline
\hline
        &$(1,3)$    &$D^*$      & $-\sqrt{\frac{1}{2}}\,g_{63}^{\bar3}\,h_{\bar39}^{\bar3} - \sqrt{\frac{1}{2}}\,g_{63}^{\bar3}\,h_{9\bar3}^{\bar3}$ \\
\cline{2-3}
        &$(2,3)$    &$D^*_s$    & $-\sqrt{\frac{1}{2}}\,g_{63}^{\bar3}\,h_{\bar39}^{\bar3} - \sqrt{\frac{1}{2}}\,g_{63}^{\bar3}\,h_{9\bar3}^{\bar3}$ \\
\cline{2-3}
        &$(3,3)$    &$\rho$     & $-\frac12\,g_{66}^9\,h_{\bar3\bar3}^9$ \\
\end{tabular}
\caption{The coupled-channel structure of the t-channel exchange in (\ref{def-K}).}
\label{tab:appendix:10}
\end{center}
\end{table}

\begin{table}
\rescale
\setlength{\tabcolsep}{1.2mm}
\setlength{\arraycolsep}{3.0mm}
\renewcommand{\arraystretch}{1.0}
\begin{center}
\begin{tabular}{|l|l|l|c|}
\hline
\makebox[1cm][l]{$(I,\phantom{x}S)$} & \makebox[1cm][l]{$(a,b)$} &
\makebox[1.cm][l]{$V$} & \makebox[9.8cm][l]{$C_{V,ab}^{(I,S,3)}\qquad \qquad \quad $ charm three} \\
\hline
\hline
$(\frac12,-1)$  &$(1,1)$    &$\omega$   & $4\,g_{11}^1\,h_{99}^9$ \\
        &       &$\phi$     & $-4\,g_{11}^1\,h_{99}^9$ \\
\cline{2-3}
        &$(1,2)$    &$D^*_s$    & $g_{31}^{\bar3}\,h_{\bar39}^{\bar3} + g_{31}^{\bar3}\,h_{9\bar3}^{\bar3}$ \\
\cline{2-3}
        &$(2,2)$    &$\omega$   & $-8\,g_{33}^1\,h_{\bar3\bar3}^1 - 4\,g_{33}^9\,h_{\bar3\bar3}^1 - 2\,g_{33}^1\,h_{\bar3\bar3}^9 - g_{33}^9\,h_{\bar3\bar3}^9$ \\
        &       &$\phi$     & $-4\,g_{33}^1\,h_{\bar3\bar3}^1 - 2\,g_{33}^1\,h_{\bar3\bar3}^9$ \\
        &       &$J/\Psi$   & $-2\,g_{33}^0\,h_{\bar3\bar3}^0$ \\
\hline
\hline
\makebox[1cm][l]{$(I,\phantom{x}S)$} & \makebox[1cm][l]{$(a,b)$} &
\makebox[1.cm][l]{$V$} & \makebox[9.8cm][l]{$C_{V,ab}^{(I,S,4)}\qquad \qquad \quad $ charm four} \\
\hline
\hline
$(0,+1)$    &$(1,1)$    &$\omega$   & $-8\,g_{11}^1\,h_{\bar3\bar3}^1 - 4\,g_{11}^1\,h_{\bar3\bar3}^9$ \\
        &       &$\phi$     & $-4\,g_{11}^1\,h_{\bar3\bar3}^1$ \\
        &       &$J/\Psi$   & $-2\,g_{11}^0\,h_{\bar3\bar3}^0$ \\
\hline
\hline
$(\frac12,+0)$  &$(1,1)$    &$\omega$   & $-8\,g_{11}^1\,h_{\bar3\bar3}^1 - 2\,g_{11}^1\,h_{\bar3\bar3}^9$ \\
        &       &$\phi$     & $-4\,g_{11}^1\,h_{\bar3\bar3}^1 - 2\,g_{11}^1\,h_{\bar3\bar3}^9$ \\
        &       &$J/\Psi$   & $-2\,g_{11}^0\,h_{\bar3\bar3}^0$ \\
\hline
\hline
\end{tabular}
\caption{The coupled-channel structure of the t-channel exchange in (\ref{def-K}).}
\label{tab:appendix:11}
\end{center}
\end{table}

\end{document}